\documentclass[aps,pre,onecolumn,floatfix]{revtex4}
\pdfoutput=1

\usepackage{graphicx}
\usepackage{amssymb,amsfonts,amsmath}
\usepackage{epsf}
\usepackage{subfigure}
\usepackage{epstopdf}
\usepackage{mathrsfs}

\newcommand{\be}{\begin{equation}}
\newcommand{\ee}{\end{equation}}
\newcommand{\bea}{\begin{eqnarray}}
\newcommand{\eea}{\end{eqnarray}}

\begin{document}

(J. Stat. Mech. P03024, 2011)

\vspace{1cm}

\title{\bf Nonlinear transport effects in mass separation by effusion}

\author{Pierre Gaspard and David Andrieux}
\affiliation{Center for Nonlinear Phenomena and Complex Systems,\\
Universit\'e Libre de Bruxelles, Code Postal 231, Campus Plaine,
B-1050 Brussels, Belgium}

\begin{abstract}
Generalizations of Onsager reciprocity relations are established for the nonlinear
response coefficients of ballistic transport in the effusion of gaseous mixtures.
These generalizations, which have been established 
on the basis of the fluctuation theorem for the currents,
are here considered for mass separation by effusion.
In this kinetic process, the mean values of the currents depend nonlinearly on the
affinities or thermodynamic forces controlling the nonequilibrium constraints.
These nonlinear transport effects are shown to play an important role
in the process of mass separation.  In particular, the entropy efficiency turns out to
be significantly larger than it would be the case if the currents were supposed
to depend linearly on the affinities.
\end{abstract}

\maketitle

\section{Introduction}

For long, the understanding of nonequilibrium processes has been limited to the linear
regime close to the thermodynamic equilibrium with results such as the Onsager reciprocity relations \cite{O31}.
Nowadays, great advances have been carried out since the discovery of time-reversal symmetry relations
for large-fluctuation properties on the basis of dynamical systems theory \cite{ECM93,GC95,TG95}.
Recently, these relations have been extended to stochastic processes, 
as well as to open quantum systems \cite{K98,LS99,EHM09,AGMT09}.
One of the most detailed versions of such relations is the fluctuation theorem for all the currents flowing
through a nonequilibrium system \cite{AG07JSP,A09}.  Thanks to this fluctuation theorem, 
relationships have been obtained that generalize the Onsager reciprocity relations 
from linear to nonlinear response coefficients \cite{AG04,AG06,AG07JSM}.
These relations have already been applied to linear chemical reactions \cite{AG08PRE}, 
to Brownian sieves and molecular machines \cite{Astu09}, and to electronic quantum transport
\cite{SU08,NYHCKOLESUG10}.

The fact is that many nanosystems typically develop highly nonlinear responses 
to multiple nonequilibrium driving forces, as it is the case in electronic transistors for instance \cite{AG06}.  
In such devices, several currents are coupled together so that different driving forces 
may induce the response of one particular current.
Since they only hold in linear regimes, the Onsager reciprocity relations are of limited use 
to study such nonlinear devices and we may expect that their recently discovered 
generalizations to nonlinear regimes would be valuable to understand the coupling 
between transport properties in far-from-equilibrium systems \cite{AG04,AG06,AG07JSM}.

The purpose of the present paper is to address this issue for the process of mass separation by effusion.
In this process, a mixture of two or more gases is forced to flow from one reservoir 
to another through a small orifice with a size smaller than the mean free path \cite{B91}.  
Effusion couples the heat and particle flows and kinetic theory has shown 
how these flows depend on the mass of the particles \cite{B91,K09,P58,P72}.
For a binary mixture, three currents are thus coupled together:  the heat flow, the total mass flow, 
and the differential flow of one species with respect to the other.
Several coupling effects are possible between these three flows and, in particular,
the separation between particles of different masses induced by either the temperature
or the pressure difference between both reservoirs.

Although a fluctuation theorem has recently been obtained for effusion \cite{CVK06,WVKL07,CWEV08}, neither its extension
to mass separation nor its consequences on the nonlinear response coefficients have been reported on.
These coefficients are worth to be studied because they obey remarkable relationships
that are the consequence of microreversibility \cite{AG04,AG06,AG07JSM}, 
as the Onsager reciprocity relations are in linear regimes \cite{O31}.
Furthermore, these coefficients control the coupling between the different currents far from equilibrium
and they influence the efficiency of mass separation.  In this context, a concept of entropy efficiency
was introduced as the ratio between the contribution to entropy production caused by mixing of the
two constituents over the one caused by heat flow \cite{O39,JF46}.  It turns out that the entropy efficiency
may be significantly higher in nonlinear regimes than in linear ones, as we show here below.

The paper is organized as follows.  In Section \ref{sec:FT}, we present the fluctuation theorem for
all the currents in the effusion of gas mixtures.  The generating function of the counting statistics of
the different flows is obtained from kinetic theory.  In Section \ref{sec:NL}, the mean currents and their
statistical cumulants are calculated from the generating function and we verify that the nonlinear
response coefficients indeed obey relationships previously reported in Refs.~\cite{AG04,AG06,AG07JSM} 
and which find their origin in microreversibility.  In Section \ref{sec:MS}, the results are applied to
the mass separation of two species.  The entropy efficiency is shown to be significantly higher
than possible if linearity was assumed.  The conclusions are drawn in Section \ref{sec:Conclusions}.

\section{Fluctuation theorem for currents in the effusion of gas mixtures}
\label{sec:FT}

\subsection{Thermodynamic formulation}

\begin{figure}[htbp]
\centerline{\includegraphics[width=8cm]{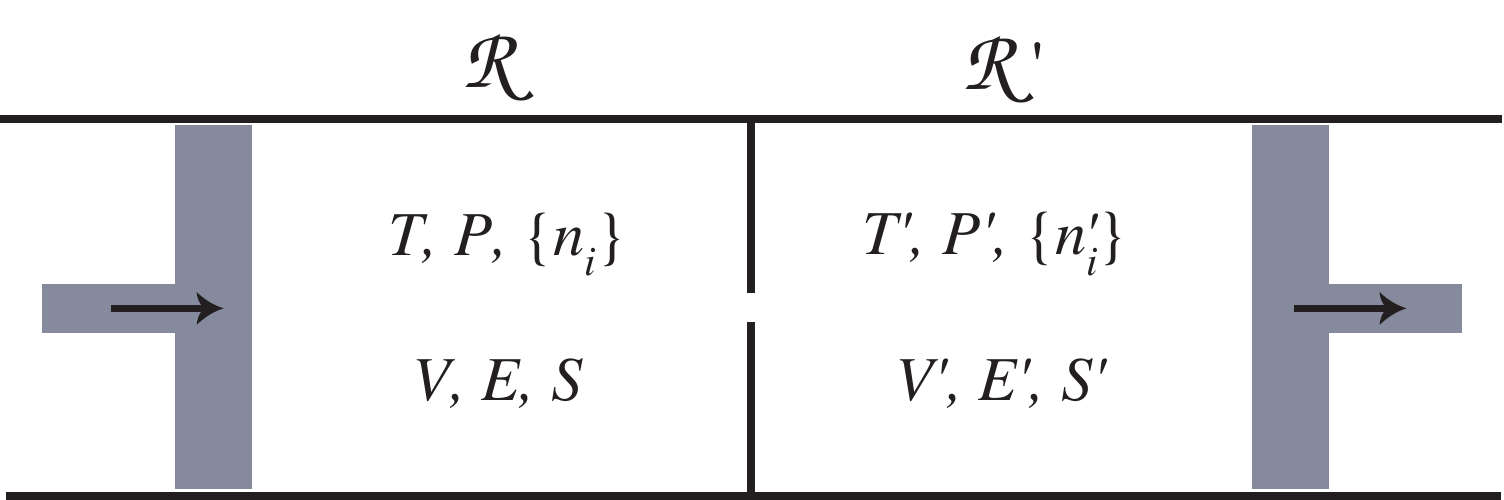}}
\caption{Schematic representation of the stationary effusion of gaseous mixtures through a small pore 
between two reservoirs $\mathcal{R}$ and $\mathcal{R}'$ where the thermodynamic conditions 
are kept fixed by moving the pistons.  These conditions are defined by the temperature $T$, the pressure $P$, 
and the particle densities $\{n_i\}_{i=1}^c$.  $V$, $E$, and $S$ denote the volume, the energy, and the entropy.  
The prime refers to the same quantities in the right reservoir.}
\label{fig1}
\end{figure}

We consider two gaseous mixtures at different temperatures, pressures and concentrations
in two reservoirs $\mathcal{R}$ and $\mathcal{R}'$ separated by a wall, through which a small
pore allows the flow of particles.  The reservoirs are assumed to be so large that the system is 
maintained out of equilibrium in a steady state.  Alternatively, the far ends of both reservoirs 
can be closed by pistons moving in such a way that the thermodynamic conditions 
of the gaseous mixtures are kept stationary (see Fig. \ref{fig1}).

The gases are supposed to be ideal mixtures of $c$ monoatomic species so that the particle densities 
$\{n_i\}_{i=1}^c$ and $\{n'_i\}_{i=1}^c$ are related to the pressures 
and temperatures by the perfect-gas equation of state:
\be
P=\sum_{i=1}^c n_i kT  \qquad\mbox{and}\qquad P'=\sum_{i=1}^c n'_i kT' \, ,
\ee
where $k$ is Boltzmann's constant.  

The currents of energy and particles from the left reservoir to the right one are defined by
\bea
&& J_E \equiv -\dot E = + \dot E' \, ,  \label{JE_dfn}\\
&& J_i \equiv -\dot N_i = + \dot N'_i \qquad (i=1,2,...,c) \, ,  \label{Ji_dfn}
\eea
in terms of the energies $E$ and $E'$, and particle numbers $N_i=n_iV$ and $N'_i=n'_iV'$ in both reservoirs.
The stationary assumption that the pressures, temperatures, and particle densities remain constant 
in both reservoirs implies that the volumes of both reservoirs should satisfy the conditions
\be
\dot V = -\frac{\sum_{i=1}^c J_i}{\sum_{i=1}^c n_i} \qquad\mbox{and}\qquad 
\dot V' = +\frac{\sum_{i=1}^c J_i}{\sum_{i=1}^c n'_i} \, ,
\ee
so that
\be
\frac{P}{T} \,\dot V + \frac{P'}{T'} \,\dot V'=0 \, .
\label{vol}
\ee

Since the system does not exchange entropy with its environment, $d_{\rm e}S/dt=0$, the time derivative 
of its total entropy $S_{\rm tot}=S+S'$ gives the entropy production of effusion 
$dS_{\rm tot}/dt=d_{\rm e}S/dt+d_{\rm i}S/dt=d_{\rm i}S/dt$.  
Using Gibbs' relation $dE=TdS-PdV+\sum_{i=1}^c\mu_idN_i$ 
for both gaseous mixtures and Eq.~(\ref{vol}), the entropy production of effusion is given by
\be
\frac{1}{k}\frac{d_{\rm i}S}{dt} = A_E\,  J_E + \sum_{i=1}^c A_i \, J_i \, ,
\label{entrprod}
\ee
in terms of the affinities or thermodynamic forces defined as \cite{DD36,P67,NP77,C85}
\bea
&& A_E \equiv \frac{1}{kT'}-\frac{1}{kT} \, , \label{AE}\\
&& A_i \equiv \frac{\mu_i}{kT}-\frac{\mu'_i}{kT'} \qquad (i=1,2,...,c) \, , \label{Ai}
\eea
where $\{\mu_i\}_{i=1}^c$ and $\{\mu'_i\}_{i=1}^c$ are the chemical potentials of the different species 
composing the gaseous mixtures in both reservoirs.  Since the gases are here supposed to be ideal 
and composed of monoatomic species, the affinities of the particle flows are given by
\be
A_i = \ln\left[ \frac{n_i}{n'_i}\left(\frac{T'}{T}\right)^{3/2}\right] \qquad (i=1,2,...,c) \, .
\ee
If we take the temperature and densities of the reservoir $\mathcal{R}$ as the reference conditions, 
the same quantities in the other reservoir $\mathcal{R}'$ can be expressed in terms of the affinities as
\bea
&& kT' = \frac{kT}{1+kT\, A_E} \, , \label{T'}\\
&& n'_i = \frac{n_i \, \exp(-A_i)}{(1+kT\, A_E)^{3/2}} \qquad (i=1,2,...,c) \, . \label{n'_i}
\eea
In this regard, the affinities are the control parameters of the nonequilibrium constraints.

Now, we still need to determine the heat and particle currents (\ref{JE_dfn})-(\ref{Ji_dfn}).  
For this purpose, kinetic theory will allow us to take into account the mechanistic aspects of effusion.

\subsection{Kinetic-theory formulation}

Effusion proceeds as a random sequence of particle passages through the pore in both ways between the two reservoirs: 
$\mathcal{R}\rightleftharpoons\mathcal{R}'$.  These random events are statistically independent of each other 
as long as the interaction between the particles is supposed to be negligible.  
In this respect, the size of the pore should be smaller 
than the mean free path of the particles in the gaseous mixture.
Therefore, the effects of binary collisions between the particles can be ignored 
and the flow of particles across the pore can be considered to be purely ballistic.
We further assume that the thickness of the wall separating both reservoirs is smaller than the diameter
of the pore so that most particles are crossing the pore in straight trajectories 
with negligible bouncing or interaction inside the pore \cite{STG09}.  

Under these conditions, the flow of particles can be determined following standard
methods from the kinetic theory of gases \cite{B91,K09,P58,P72,CVK06}.
The mean number of particles passing the cross-section area $\sigma$ of the pore
during some time interval can be calculated in terms of the Maxwellian velocity distribution
and the particle density in the reservoir, from which the particles arrive.
Since each particle of velocity $\bf v$ carries the kinetic energy $\epsilon=m_i{\bf v}^2/2$, 
the flow of particles also determines the heat flow.  
Kinetic theory describes the transport of energy and particles 
between the reservoirs as a Markovian stochastic process \cite{B91,K09,P58,P72,CVK06}.
The probability $\mathcal{P}_t(\Delta E,\Delta{\bf N})$ that an energy $\Delta E$ 
and $\Delta{\bf N}=\{\Delta N_j\}_{j=1}^c$ particles are transferred 
in the direction $\mathcal{R}\to\mathcal{R}'$ during the time interval $t$ is ruled by the master equation:
\bea
\frac{d}{dt}\mathcal{P}_t(\Delta E,\Delta{\bf N}) &=& \sum_{i=1}^c\int_0^{\infty} d\epsilon\, 
W_i^{(+)}(\epsilon) \, \mathcal{P}_t(\Delta E-\epsilon,\Delta{\bf N}-{\bf 1}_i) \nonumber\\
&& + \sum_{i=1}^c\int_0^{\infty} d\epsilon\, W_i^{(-)}(\epsilon) \, \mathcal{P}_t(\Delta E+\epsilon,\Delta{\bf N}+{\bf 1}_i)\nonumber\\
&& - \mathcal{P}_t(\Delta E,\Delta{\bf N}) \sum_{i=1}^c\int_0^{\infty} d\epsilon \left[ W_i^{(+)}(\epsilon) +W_i^{(-)}(\epsilon)\right] \, ,
\label{master}
\eea
where we use the notation ${\bf 1}_i=\{\delta_{ij}\}_{j=1}^c$ with the Kronecker symbol $\delta_{ij}$ 
and where the transition rates are given by
\bea
W_i^{(+)}(\epsilon) &=& 
\frac{\sigma\, n_i}{\sqrt{2\pi m_i kT}} \, \frac{\epsilon}{kT} \, \exp\left(-\frac{\epsilon}{kT}\right) \, , \label{rate+}\\
W_i^{(-)}(\epsilon) &=& 
\frac{\sigma\, n'_i}{\sqrt{2\pi m_i kT'}} \, \frac{\epsilon}{kT'} \, \exp\left(-\frac{\epsilon}{kT'}\right)\, , \label{rate-}
\eea
for the particle flows $\mathcal{R}\rightarrow\mathcal{R}'$ and $\mathcal{R}\leftarrow\mathcal{R}'$ 
respectively \cite{B91,K09,P58,P72,CVK06}.
We notice that these transition rates are related to each other and to the affinities (\ref{AE})-(\ref{Ai}) by
\be
\frac{W_i^{(+)}(\epsilon)}{W_i^{(-)}(\epsilon)} = \exp\left( \epsilon A_E + A_i \right) \qquad (i=1,2,...,c) \, , \label{W-ratio}
\ee
as the consequence of the nonequilibrium constraints on the flow of particles between both reservoirs.
At the equilibrium thermodynamic state where the affinities vanish, the two transition rates are equal
and the conditions of detailed balancing are recovered.  The relations (\ref{W-ratio}) are the analogues
for the present system of Schnakenberg's conditions defining the macroscopic affinities driving general
nonequilibrium stochastic processes \cite{AG07JSP,A09,AG04,AG06,S76}.

We introduce the generating function of the currents and their statistical cumulants as
\be
Q(\lambda_E,\pmb{\lambda}) \equiv \lim_{t\to\infty} -\frac{1}{t}
 \ln \left\langle \exp\left(-\lambda_E\, \Delta E-\pmb{\lambda}\cdot\Delta{\bf N}\right)\right\rangle_t \, ,
 \label{G_dfn}
\ee
where the statistical average is defined by
\be
\left\langle \exp\left(-\lambda_E\, \Delta E-\pmb{\lambda}\cdot\Delta{\bf N}\right)\right\rangle_t 
= \sum_{\Delta E,\Delta{\bf N}} \mathcal{P}_t(\Delta E,\Delta{\bf N}) \, \exp\left(-\lambda_E\, \Delta E-\pmb{\lambda}\cdot\Delta{\bf N}\right) \, ,
\ee
in terms of the probability ruled by the master equation (\ref{master}) \cite{AG04,AG06}.  
Taking the time derivative of this average and using the master equation, 
we find in the long-time limit that the generating function is given by
\be
Q(\lambda_E,\pmb{\lambda}) = \sum_{i=1}^c\int_0^{\infty} d\epsilon\, 
\left[ W_i^{(+)}(\epsilon)  \left( 1 - {\rm e}^{-\lambda_E  \epsilon-\lambda_i}\right) 
+ W_i^{(-)}(\epsilon)  \left( 1 - {\rm e}^{\lambda_E  \epsilon+\lambda_i}\right)\right] \, .
\ee
Using the transition rates (\ref{rate+}) and (\ref{rate-}), 
we get the explicit expression of the generating function:
\be
Q(\lambda_E,\pmb{\lambda}) = \sum_{i=1}^c 
\left\{ \sigma \, n_i \sqrt{\frac{kT}{2\pi m_i}} \left[ 1 - \frac{{\rm e}^{-\lambda_i}}{(1+kT\, \lambda_E)^2} \right] 
+ \sigma \, n'_i \sqrt{\frac{kT'}{2\pi m_i}} \left[ 1 - \frac{{\rm e}^{\lambda_i}}{(1-kT'\, \lambda_E)^2} \right] \right\} \, ,
\label{G}
\ee
under the condition that
\be
-\frac{1}{kT} < \lambda_E < \frac{1}{kT'} \, .
\label{condition}
\ee
The expression (\ref{G}), which was obtained for a gas with a single constituent in Ref.~\cite{CVK06}, is here extended to gaseous mixtures.

The rates of effusion into vacuum are known to be given by
\be
r_i \equiv  \sigma n_i \sqrt{\frac{kT}{2\pi m_i}} = \frac{1}{4} \sigma n_i \langle v_i \rangle \, ,
\label{rate}
\ee
where
\be
\langle v_i \rangle = \langle \Vert{\bf v}_i\Vert \rangle = \sqrt{\frac{8kT}{\pi m_i}}
\ee
is the mean value of the speed of particles of mass $m_i$ \cite{B91,K09,P58,P72}.
If we use these rates and express the temperature $T'$ and the densities $\{n'_i\}$ 
in terms of the affinities by using Eqs.~(\ref{T'})-(\ref{n'_i}),
the generating function can be rewritten as
\be
Q(\lambda_E,\pmb{\lambda}) = \sum_{i=1}^c 
r_i  \left\{ 1 + \frac{{\rm e}^{-A_i}}{(1+kT\, A_E)^2}- \frac{{\rm e}^{-\lambda_i}}{(1+kT\, \lambda_E)^2}  - \frac{{\rm e}^{-(A_i-\lambda_i)}}{\left[1+kT(A_E-\lambda_E)\right]^2} \right\} \, .
\label{Gbis}
\ee

Consequently, we observe that the generating function satisfies the following symmetry relation:
\be
Q(\lambda_E,\pmb{\lambda}) = Q(A_E-\lambda_E,{\bf A}-\pmb{\lambda})  \, ,
\label{FT}
\ee
which is the {\it fluctuation theorem for the currents} \cite{AG07JSP,A09}.
We notice that the generating function vanishes according to
\be
Q(0,{\bf 0}) = Q(A_E,{\bf A}) = 0  \, ,
\label{norm}
\ee
because of its definition (\ref{G_dfn}), 
the normalization condition $\langle 1 \rangle_t=1$, 
and the symmetry relation (\ref{FT}).
An equivalent expression of the fluctuation theorem for the currents is given by
\be
\frac{{\mathcal P}_t(\Delta E, \Delta{\bf N})}{{\mathcal P}_t(-\Delta E, -\Delta{\bf N})} 
\simeq \exp(A_E\,  \Delta E + {\bf A}\cdot\Delta{\bf N}) \qquad\mbox{for} \quad t\to\infty \, .
\label{FT2}
\ee
In this form, the fluctuation theorem compares the probabilities of opposite fluctuations 
in the effusive transport of energy and particles across the pore
and shows that the ratio of these probabilities is controlled by the affinities.  
At the thermodynamic equilibrium where the affinities are vanishing,
the principle of detailed balancing is recovered according to which the 
probabilities of opposite fluctuations are in balance.
Out of equilibrium, this balance is broken by the nonequilibrium probability distribution 
and the fluctuating currents tend to follow the directionality fixed by the affinities.
The fluctuation theorem (\ref{FT2}) is the consequence of the relations (\ref{W-ratio})
satisfied by the transition rates of particles of energy $\epsilon$ and species $i=1,2,...,c$
between both reservoirs.  The difference between Eqs. (\ref{W-ratio}) and (\ref{FT2})
is that the relations (\ref{W-ratio}) concern the rates of random transfers 
of individual particles over an infinitesimal time interval, although the fluctuation
theorem (\ref{FT2}) holds for the cumulative transport of an energy $\Delta E$ 
and particle numbers $\Delta{\bf N}$ over the long time interval $t$.
Otherwise, the latter results from the former in much the same way
as the fluctuation theorem for currents can be deduced from Schnakenberg's
definition of macroscopic affinities in general nonequilibrium stochastic processes
\cite{AG07JSP,A09,AG04,AG06,S76}.

Now, the average values of the currents can be deduced from the generating function (\ref{Gbis}) according to
\bea
J_E &=& \frac{\partial Q}{\partial\lambda_E}(0,{\bf 0}) = \lim_{t\to\infty} \frac{1}{t} \langle\Delta E\rangle_t \, ,\\
J_i &=& \frac{\partial Q}{\partial\lambda_i}(0,{\bf 0}) = \lim_{t\to\infty} \frac{1}{t} \langle\Delta N_{i}\rangle_t \qquad (i=1,2,...,c)\, .
\eea
The energy and particle currents are thus given in terms of the affinities according to
\bea
J_E &=& 2 \, kT \, \sum_{i=1}^c  r_i \left[ 1 - \frac{{\rm e}^{-A_i}}{(1+kT\, A_E)^3}\right]  \, ,  \label{JE}\\
J_i &=& r_i \, \left[ 1 - \frac{{\rm e}^{-A_i}}{(1+kT\, A_E)^2}\right]  \qquad (i=1,2,...,c)\, . \label{Ji}
\eea
We recover the well-known result that, if there is the vacuum in the reservoir $\mathcal{R}'$, 
the energy and particle currents become
\bea
\lim_{{\bf A}\to\infty} J_E &=& 2 \, kT \, \sum_{i=1}^c r_i \, ,  \\
\lim_{{\bf A}\to\infty} J_i &=& r_i  \qquad \qquad\qquad (i=1,2,...,c)\, 
\eea
where $r_i$ is the effusion rate given by Eq.~(\ref{rate}) \cite{B91,K09,P58,P72,CVK06}.

Using the inequality $\left\langle{\rm e}^{-X}\right\rangle \geq {\rm e}^{-\langle X\rangle}$ \cite{inequality}, we have that
\be
\left\langle{\rm e}^{-A_E\,  \Delta E - {\bf A}\cdot\Delta{\bf N}}\right\rangle_t \geq {\rm e}^{-A_E\,  \langle\Delta E\rangle_t - {\bf A}\cdot\langle\Delta{\bf N}\rangle_t}  \, .
\label{inequal}
\ee
Therefore, in the macroscopic steady state reached in the long-time limit, we infer from the inequality~(\ref{inequal}) and Eq.~(\ref{norm}) that
\bea
\frac{1}{k} \frac{d_{\rm i}S}{dt} &=& A_E\, J_E+{\bf A}\cdot{\bf J} 
= \lim_{t\to\infty} \frac{1}{t} \left(A_E\,  \langle\Delta E\rangle_t + {\bf A}\cdot\langle\Delta{\bf N}\rangle_t\right) \nonumber\\
&\geq& \lim_{t\to\infty} - \frac{1}{t} \ln \left\langle{\rm e}^{-A_E\,  \Delta E - {\bf A}\cdot\Delta{\bf N}}\right\rangle_t = Q(A_E,{\bf A})=0 \, ,
\label{2ndlaw}
\eea
which is the verification that the second law of thermodynamics holds in the present formulation.

\section{Nonlinear response coefficients}
\label{sec:NL}

The authors of the present paper have previously proved that the symmetry relation (\ref{FT})
has for consequence not only the Onsager reciprocity relations which only hold in the linear regimes,
but also generalizations of these relations 
which extend to the nonlinear response coefficients \cite{AG04,AG06,AG07JSM,G10}.
These generalizations of Onsager reciprocity relations are remarkable,
in particular, because they can be used to express the response coefficients 
in terms of quantities characterizing the diffusivities of the currents and higher cumulants
in the regime of nonlinear transport by effusion.

In this section, we use the following notations:
\bea
&& A=\{A_{\alpha}\}=\{A_E,{\bf A}\} = \{A_E,A_1,A_2,...,A_c\} \, , \\
&& \lambda=\{\lambda_{\alpha}\}=\{\lambda_E,\pmb{\lambda}\} = \{\lambda_E,\lambda_1,\lambda_2,...,\lambda_c\} \, ,
\eea
with the indices $\alpha\in\{E,1,2,...,c\}$.

The response coefficients $L_{\alpha,\beta}$, $M_{\alpha,\beta\gamma}$, $N_{\alpha,\beta\gamma\delta}$, etc... are defined by expanding the average values of the currents in powers of the affinities:
\be
J_\alpha(A)=\frac{\partial Q}{\partial\lambda_\alpha}\Big\vert_{\lambda=0}= \sum_\beta L_{\alpha,\beta}A_\beta
+\frac{1}{2}\sum_{\beta,\gamma} M_{\alpha,\beta\gamma}A_\beta A_\gamma 
+\frac{1}{6}\sum_{\beta,\gamma,\delta} N_{\alpha,\beta\gamma\delta}A_\beta A_\gamma A_\delta + \cdots
\ee
Here, the comma separates the index of the current from the indices of the affinities.
We notice that the tensors $M_{\alpha,\beta\gamma}$, $N_{\alpha,\beta\gamma\delta}$,... are {\it a priori} only symmetric under the exchange of the indices beyond the comma.

On the other hand, the diffusivities and the higher cumulants of the fluctuating currents are defined by successive derivatives of the generating function with respect to the parameters $\lambda$:
\bea
&& D_{\alpha\beta}(A) \equiv - \frac{1}{2} \frac{\partial^2Q}{\partial\lambda_\alpha\partial\lambda_\beta}\Big\vert_{\lambda=0} \, , \label{2nd_cumul}\\
&& C_{\alpha\beta\gamma}(A) \equiv \frac{\partial^3Q}{\partial\lambda_\alpha\partial\lambda_\beta\partial\lambda_\gamma}\Big\vert_{\lambda=0} \, , \label{3rd_cumul}\\
&& B_{\alpha\beta\gamma\delta}(A) \equiv - \frac{1}{2} \frac{\partial^4Q}{\partial\lambda_\alpha\partial\lambda_\beta\partial\lambda_\gamma\partial\lambda_\delta}\Big\vert_{\lambda=0} \, , \label{4th_cumul}\\
&& \qquad\qquad\qquad\qquad\vdots \nonumber
\eea
so that all these tensors are totally symmetric.

The fluctuation theorem for the currents (\ref{FT}) has for consequence the following realtionships starting from the Onsager reciprocity relations for the linear response coefficients and extending to the nonlinear response coefficients \cite{AG04,AG06,AG07JSM,G10}
\bea
&& L_{\alpha,\beta}=D_{\alpha\beta}(0)=L_{\beta,\alpha} \, , \label{L}\\
&& M_{\alpha,\beta\gamma} = \left( \frac{\partial D_{\alpha\beta}}{\partial A_\gamma} + \frac{\partial D_{\alpha\gamma}}{\partial A_\beta} \right)_{A=0} \, , \label{M}\\
&& N_{\alpha,\beta\gamma\delta} = \left( \frac{\partial^2 D_{\alpha\beta}}{\partial A_\gamma\partial A_\delta} + \frac{\partial^2 D_{\alpha\gamma}}{\partial A_\beta\partial A_\delta} +\frac{\partial^2 D_{\alpha\delta}}{\partial A_\beta\partial A_\gamma} -\frac{1}{2} B_{\alpha\beta\gamma\delta}\right)_{A=0} \, , \label{N}\\
&& \qquad\qquad\qquad\qquad\vdots \nonumber
\eea
A generalization of Onsager reciprocity relations is given by the total symmetry of the following fourth-order tensor:
\be
N_{\alpha,\beta\gamma\delta} - \left( \frac{\partial^2 D_{\alpha\beta}}{\partial A_\gamma\partial A_\delta} + \frac{\partial^2 D_{\alpha\gamma}}{\partial A_\beta\partial A_\delta} +\frac{\partial^2 D_{\alpha\delta}}{\partial A_\beta\partial A_\gamma}\right)_{A=0} \, , \label{symN}
\ee
which is the consequence of Eq. (\ref{N}) and the total symmetry 
of the fourth-cumulant tensor (\ref{4th_cumul}) \cite{AG04,AG06,AG07JSM}.
Moreover, the fluctuation theorem for the currents (\ref{FT}) has also for consequence the following relations \cite{AG07JSM}
\be
B_{\alpha\beta\gamma\delta}(0) = \left( \frac{\partial C_{\alpha\beta\gamma}}{\partial A_\delta} \right)_{A=0} \, .
\label{B}
\ee

Our purpose is now to apply these relationships to the effusion process characterized by the generating function (\ref{Gbis}).

Because the currents are fluctuating, the energy $\Delta E$ and the particle numbers $\Delta {\bf N}$ transferred over a time interval $t$ undergo random walks around their mean values, which are characterized by the diffusivities or second-order cumulants (\ref{2nd_cumul}):
\bea
&& D_{EE} = 3 \, (kT)^2\sum_{i=1}^c r_i \left[ 1 + \frac{{\rm e}^{-A_i}}{(1+ kT \, A_E)^4} \right] \, , \\
&& D_{Ei} =  \ kT \, r_i \left[ 1 + \frac{{\rm e}^{-A_i}}{(1+ kT \, A_E)^3} \right] \, , \\
&& D_{ii} =  \frac{1}{2}\, r_i  \left[ 1 + \frac{{\rm e}^{-A_i}}{(1+ kT \, A_E)^2} \right] \, , 
\eea
while $D_{ij}=0$ for $i\neq j$.
Among the large-deviation properties, we also have the third-order cumulants (\ref{3rd_cumul}):
\bea
&& C_{EEE} = 24 \, (kT)^3\sum_{i=1}^c r_i \left[ 1 - \frac{{\rm e}^{-A_i}}{(1+ kT \, A_E)^5} \right] \, , \\
&& C_{EEi} =  6 \, (kT)^2 \, r_i \left[ 1 - \frac{{\rm e}^{-A_i}}{(1+ kT \, A_E)^4} \right] \, , \\
&& C_{Eii} =  2 \, kT \, r_i  \left[ 1 - \frac{{\rm e}^{-A_i}}{(1+ kT \, A_E)^3} \right] \, , \\
&& C_{iii} =  \  r_i  \left[ 1 - \frac{{\rm e}^{-A_i}}{(1+ kT \, A_E)^2} \right] \, , 
\eea
while the other third-order cumulants are vanishing.  The fourth-order cumulants can be obtained 
by using their definition (\ref{4th_cumul}) in terms of the generating function or
thanks to the relations (\ref{B}):
\bea
B_{EEEE}(0) =&\displaystyle   \frac{\partial C_{EEE}}{\partial A_E}(0) &= 120 \,(kT)^4  \sum_{i=1}^c r_i \, , \label{BEEEE}\\
B_{EEEi}(0) =&\displaystyle  \frac{\partial C_{EEE}}{\partial A_i}(0)=\frac{\partial C_{EEi}}{\partial A_E}(0)   &= 24\, (kT)^3 \, r_i \, , \label{BEEEi}\\
B_{EEii}(0) =&\displaystyle  \frac{\partial C_{EEi}}{\partial A_i}(0)=\frac{\partial C_{Eii}}{\partial A_E}(0)&= 6 \, (kT)^2 \, r_i \, , \label{BEEii}\\
B_{Eiii}(0) =&\displaystyle  \frac{\partial C_{Eii}}{\partial A_i}(0)=\frac{\partial C_{iii}}{\partial A_E}(0)&= 2 \, kT\, r_i \, , \label{BEiii}\\
B_{iiii}(0) =&\displaystyle  \frac{\partial C_{iii}}{\partial A_i}(0)&=  \ r_i \, , \label{Biiii}
\eea
while the other cumulants are vanishing.

The remarkable relationships (\ref{L})-(\ref{N}) for the response coefficients of the currents (\ref{JE}) and (\ref{Ji}) are satisfied as well:
\bea
L_{E,E} =& D_{EE}(0) &= 6 \,(kT)^2 \, \sum_{i=1}^c r_i \, , \label{LEE}\\
L_{E,i} =& D_{Ei}(0) &= 2 \, kT \, r_i \, , \label{LEi} \\
L_{i,E} =& D_{Ei}(0) &= 2 \, kT \, r_i \, , \label{LiE} \\
L_{i,i} =& D_{ii}(0) &= \ r_i \, , \label{Lii} 
\eea
\bea
M_{E,EE} =&\displaystyle  2\, \frac{\partial D_{EE}}{\partial A_E}(0) &= -24 \,(kT)^3 \, \sum_{i=1}^c r_i \, , \\
M_{E,Ei} =&\displaystyle  \frac{\partial D_{EE}}{\partial A_i}(0)+\frac{\partial D_{Ei}}{\partial A_E}(0) &= -6 \,(kT)^2 \, r_i \, , \label{MEEi}\\
M_{i,EE} =&\displaystyle 2 \, \frac{\partial D_{Ei}}{\partial A_E}(0) &= -6 \,(kT)^2 \, r_i \, , \label{MiEE}\\
M_{E,ii} =&\displaystyle 2 \, \frac{\partial D_{Ei}}{\partial A_i}(0) &= -2\, kT \ r_i  \, , \\
M_{i,iE} =&\displaystyle \frac{\partial D_{ii}}{\partial A_E}(0)+\frac{\partial D_{Ei}}{\partial A_i}(0) &= -2\, kT \ r_i  \, , \\
M_{i,ii} =&\displaystyle  2 \, \frac{\partial D_{ii}}{\partial A_i}(0) &= \ \ - r_i \, , 
\eea
\bea
N_{E,EEE} =&\displaystyle 3\, \frac{\partial^2 D_{EE}}{\partial A_E^2}(0) - \frac{1}{2} \, B_{EEEE}(0)  &= 120 \,(kT)^4 \, \sum_{i=1}^c r_i \, , \\
N_{E,EEi} = &\displaystyle 2\, \frac{\partial^2 D_{EE}}{\partial A_E \partial A_i}(0)+ \frac{\partial^2 D_{Ei}}{\partial A_E^2}(0) - \frac{1}{2} \, B_{EEEi}(0) &= 24 \,(kT)^3 \, r_i \, , \\
N_{i,EEE} =&\displaystyle 3\, \frac{\partial^2 D_{Ei}}{\partial A_E^2}(0) - \frac{1}{2} \, B_{EEEi}(0)  &= 24 \,(kT)^3 \, r_i \, , \\
N_{E,Eii}  =&\displaystyle \frac{\partial^2 D_{EE}}{\partial A_i^2}(0) + 2\, \frac{\partial^2 D_{Ei}}{\partial A_E \partial A_i}(0) - \frac{1}{2} \, B_{EEii}(0) &= 6 \,(kT)^2 \, r_i \, , \\
N_{i,iEE} =&\displaystyle \frac{\partial^2 D_{ii}}{\partial A_E^2}(0) + 2\, \frac{\partial^2 D_{Ei}}{\partial A_E \partial A_i}(0) - \frac{1}{2} \, B_{EEii}(0) &= 6 \,(kT)^2 \, r_i \, , \\
N_{E,iii}  =&\displaystyle 3 \, \frac{\partial^2 D_{Ei}}{\partial A_i^2}(0) -\frac{1}{2} \, B_{Eiii}(0)&= 2 \, kT \, r_i  \, , \label{NEiii}\\
N_{i,iiE} =&\displaystyle 2 \, \frac{\partial^2 D_{ii}}{\partial A_E\partial A_i}(0)+\frac{\partial^2 D_{Ei}}{\partial A_i^2}(0) -\frac{1}{2} \, B_{Eiii}(0)  &= 2 \, kT \, r_i  \, , \label{NiiiE}\\
N_{iiii} =&\displaystyle 3 \, \frac{\partial^2 D_{ii}}{\partial A_i^2}(0) -\frac{1}{2} \, B_{iiii}(0) &= \  \ r_i \, ,
\eea
which shows that the response coefficients can indeed be obtained from the cumulants.

The equality between the coefficients (\ref{LEi}) and (\ref{LiE}) is the Onsager reciprocity relation $L_{E,i} =L_{i,E}$, 
which results from the total symmetry of the diffusivity tensor (\ref{2nd_cumul}). 
However, we point out that, for instance, the equalities between the coefficients (\ref{MEEi})~and~(\ref{MiEE}) [or between the coefficients (\ref{NEiii})~and~(\ref{NiiiE})] are specific
to the effusion process and they are not the consequences of Eqs. (\ref{M})-(\ref{B}) we are here concerned with. A system where such system-specific equalities do not hold has been described in Ref.~\cite{AG08PRE}.  Nevertheless, the equalities
\bea
N_{E,EEi} - 2 \, \frac{\partial^2 D_{EE}}{\partial A_E\partial A_i}(0) =&\displaystyle  N_{i,EEE} - 2 \, \frac{\partial^2 D_{Ei}}{\partial A_E^2}(0) &= 0  \, , \label{symNEEEi} \\
N_{E,Eii} - \frac{\partial^2 D_{EE}}{\partial A_i^2}(0) =&\displaystyle  N_{i,iEE} - \frac{\partial^2 D_{ii}}{\partial A_E^2}(0) =& 3 \, (kT)^2 \,  r_i  \, , \label{symNEEii} \\
N_{E,iii} - 2 \, \frac{\partial^2 D_{Ei}}{\partial A_i^2}(0) =&\displaystyle  N_{i,iiE} - 2 \, \frac{\partial^2 D_{ii}}{\partial A_E\partial A_i}(0) &= 0  \, , \label{symNEiii}
\eea
do result from the total symmetry of the fourth-order tensor (\ref{symN}) and are the mentioned generalizations of Onsager reciprocity relations  \cite{AG04,AG06,AG07JSM}.  These equalities, as well as Eqs. (\ref{BEEEE})-(\ref{Biiii}), are the consequences of the fluctuation theorem (\ref{FT}). We have thus shown that the fluctuation theorem for currents has fundamental implications on the nonlinear response coefficients, which indeed satisfy the remarkable relationships (\ref{M})-(\ref{B}) in the effusion process.

These results hold as well for other effusion processes \cite{WVKL07,CWEV08}.

\section{Mass separation}
\label{sec:MS}

\subsection{The coupling between the currents}

Here, we consider a binary mixture of particles with different masses $m_1 \neq m_2$.  Mass separation is the process in which the molar fraction of one of the species increases as the gaseous mixture flows between the reservoirs.  Since the mixing entropy should be reduced, mass separation requires energy supply.  The effusion process provides this energy by the coupling of the particle flows to the heat flow, as shown in previous sections.

For the effusion of a binary mixture, there exist three coupled flows: the heat flow and the two flows of particles of masses $m_1$ and $m_2$.  In order to achieve mass separation, the difference between the two particle flows must be controlled by the other flows.  Therefore, we introduce the current of the total number of particles and the difference between the particle currents we shall call the {\it differential current} \cite{footnote}
\bea
&& J_N \equiv J_1 + J_2 \, , \\
&& J_D \equiv J_1 -J_2 \, .
\eea
The conjugated affinities are defined as
\bea
&& A_N \equiv \frac{1}{2}(A_1 + A_2) = \frac{1}{2} \ln \left[\frac{n_1n_2}{n'_1n'_2}\left(\frac{T'}{T}\right)^3\right] \, , \label{AN}\\
&& A_D \equiv \frac{1}{2}(A_1 - A_2) = \frac{1}{2} \ln \frac{n_1/n_2}{n'_1/n'_2}\, . \label{AD}
\eea
The affinity associated with the differential current is related by $A_D=\ln(1/\sqrt{f})$ to the separation factor:
\be
f \equiv \frac{n'_1/n'_2}{n_1/n_2} \, ,
\label{f}
\ee
for the enrichment of the binary mixture by species 1 during the transfer $\mathcal{R}\to\mathcal{R}'$.

The currents are given in terms of the affinities by
\bea
J_E &=& 2 \, kT \, r_1 \left[ 1 - \frac{{\rm e}^{-A_N-A_D}}{(1+kT\, A_E)^3}\right] 
+ 2 \, kT \, r_2 \left[ 1 - \frac{{\rm e}^{-A_N+A_D}}{(1+kT\, A_E)^3}\right]  \, ,  \label{JES}\\
J_N &=& r_1 \left[ 1 - \frac{{\rm e}^{-A_N-A_D}}{(1+kT\, A_E)^2}\right] 
+ r_2 \left[ 1 - \frac{{\rm e}^{-A_N+A_D}}{(1+kT\, A_E)^2}\right] \, , \label{JN} \\
J_D &=& r_1 \left[ 1 - \frac{{\rm e}^{-A_N-A_D}}{(1+kT\, A_E)^2}\right] 
- r_2 \left[ 1 - \frac{{\rm e}^{-A_N+A_D}}{(1+kT\, A_E)^2}\right] \, . \label{JD}
\eea
If we introduce the molar fractions, also called the molar concentrations,
\be
c_i \equiv \frac{n_i}{n_1+n_2} \qquad (i=1,2) \, , 
\ee
the currents can be expressed as follows
\bea
J_E &=& 2 \, \sigma \left[ \left( \frac{c_1}{\sqrt{2\pi m_1}} + \frac{c_2}{\sqrt{2\pi m_2}}\right) P\sqrt{kT} -
\left( \frac{c'_1}{\sqrt{2\pi m_1}} + \frac{c'_2}{\sqrt{2\pi m_2}}\right) P'\sqrt{kT'} \right] \, ,  \label{JESbis}\\
J_N &=& \sigma \left[ \left( \frac{c_1}{\sqrt{2\pi m_1}} + \frac{c_2}{\sqrt{2\pi m_2}}\right) \frac{P}{\sqrt{kT}} -
\left( \frac{c'_1}{\sqrt{2\pi m_1}} + \frac{c'_2}{\sqrt{2\pi m_2}}\right) \frac{P'}{\sqrt{kT'}} \right]\, , \label{JNbis} \\
J_D &=& \sigma \left[ \left( \frac{c_1}{\sqrt{2\pi m_1}} - \frac{c_2}{\sqrt{2\pi m_2}}\right) \frac{P}{\sqrt{kT}} -
\left( \frac{c'_1}{\sqrt{2\pi m_1}} - \frac{c'_2}{\sqrt{2\pi m_2}}\right) \frac{P'}{\sqrt{kT'}} \right]\, , \label{JDbis}
\eea
in terms of the pressures and temperatures in both reservoirs.
These expressions show that all the currents $\{J_E,J_N,J_D\}$ are coupled together so that the differential current $J_D$ can be controlled by both the currents $J_E$ and $J_N$ and, therefore, by the differences of pressures or temperatures.  

\begin{figure}[htbp]
\begin{tabular}{ccccc}
  \rotatebox{0}{\scalebox{0.34}{\includegraphics{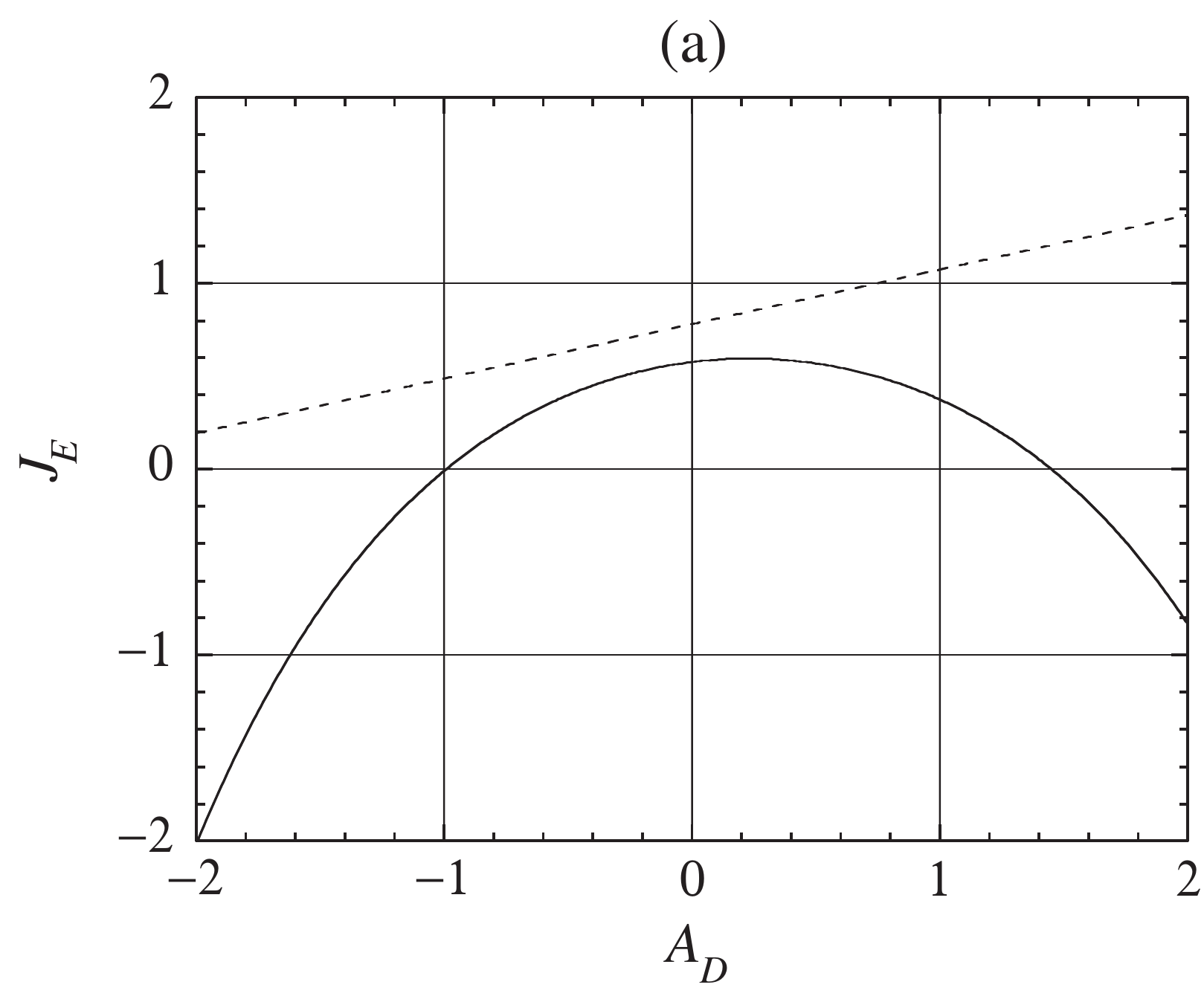}}}
  &&
  \rotatebox{0}{\scalebox{0.34}{\includegraphics{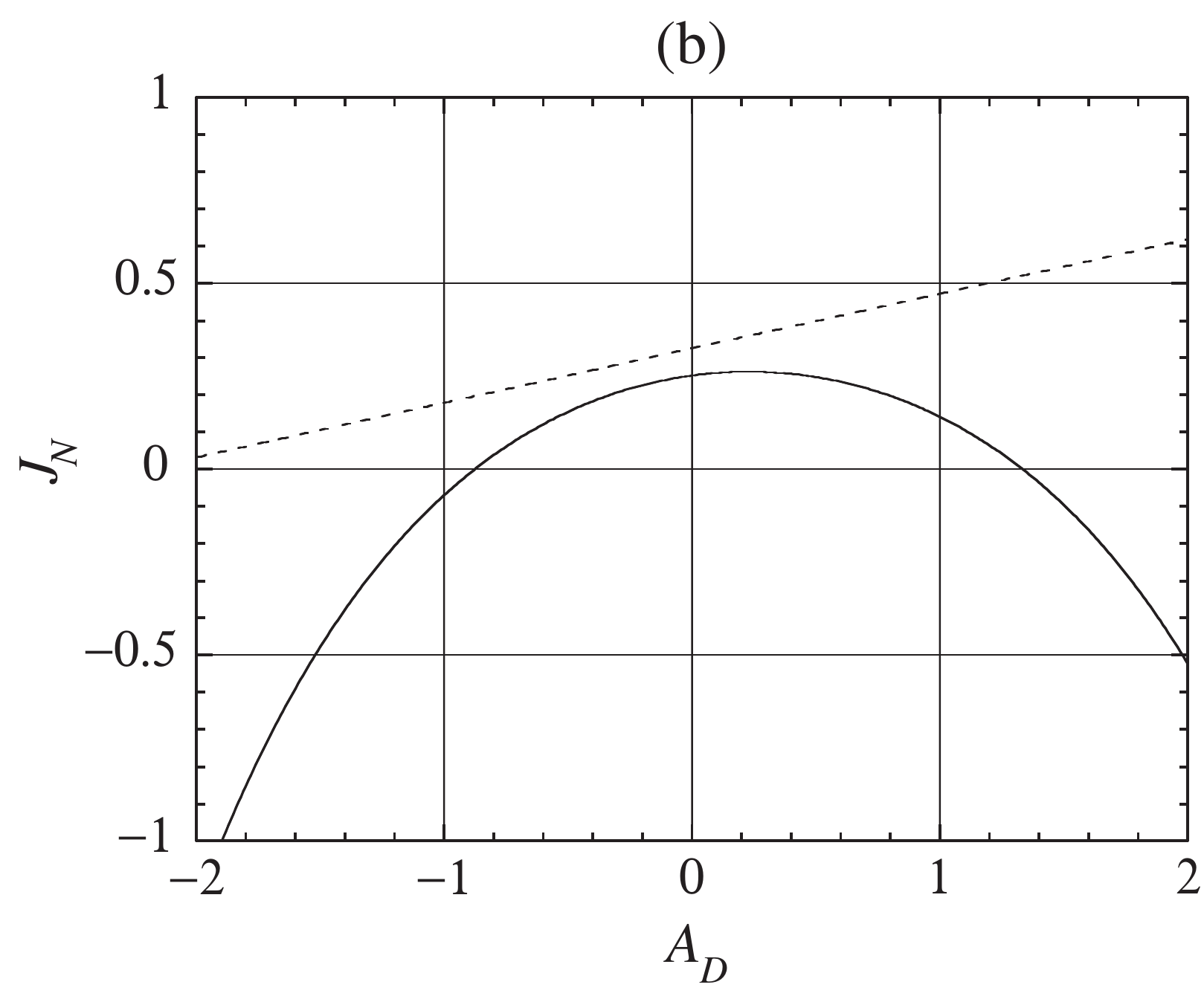}}} 
  &&
  \rotatebox{0}{\scalebox{0.34}{\includegraphics{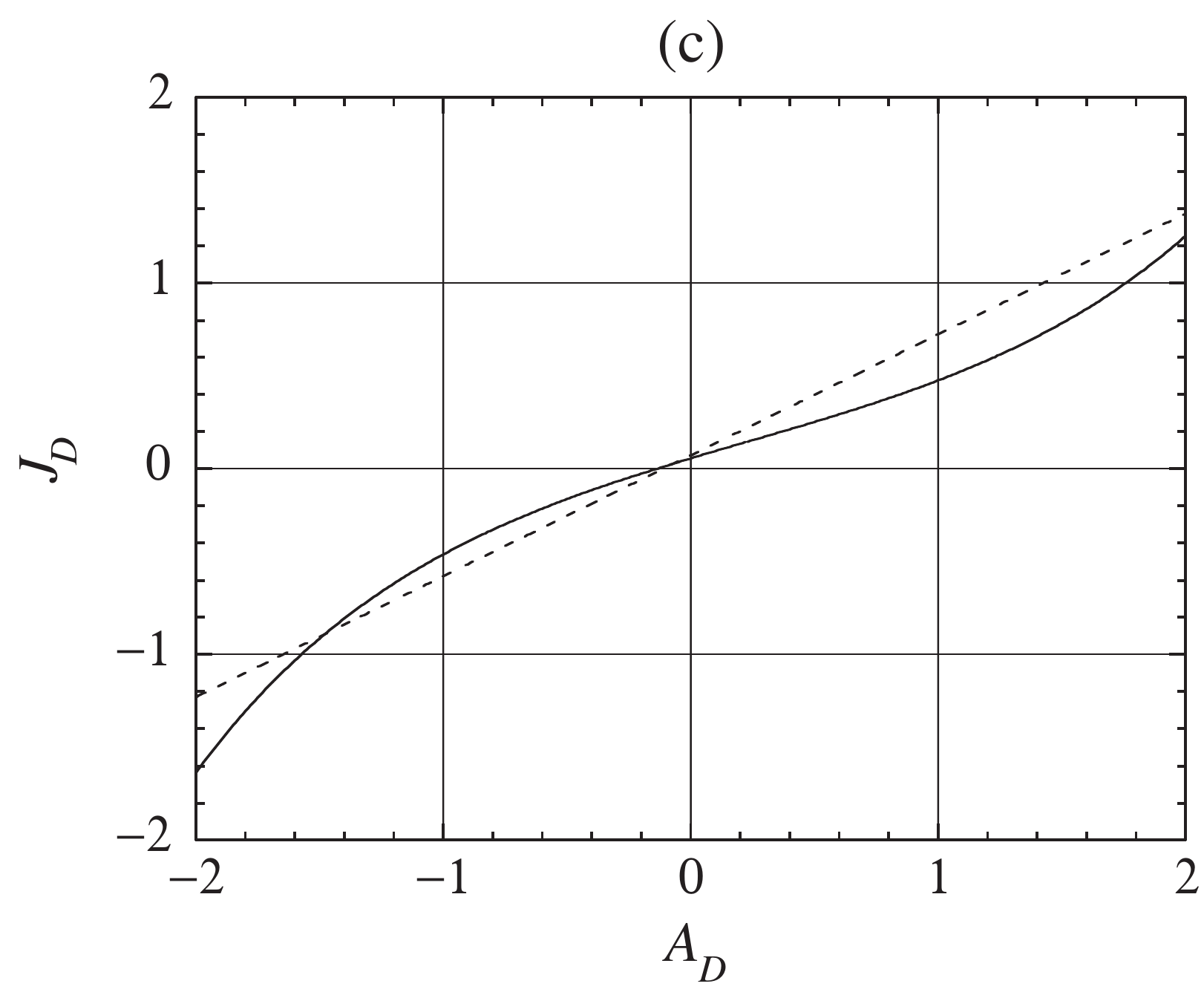}}} \\
\end{tabular}
\caption{The average currents versus the affinity $A_D$ in the effusion of 
particles of masses $m_1=1$ and $m_2=10$
composing a binary mixture at the temperature $kT=1$ and densities
$n_1=1$ and $n_2=2$ through a small pore of area $\sigma=1$ 
from the reservoir $\mathcal{R}$ to the reservoir $\mathcal{R}'$ 
at the affinities $A_E=0.1$ and $A_N=0.3$ with respect to the former:
(a) the energy current $J_E$; (b) the total particle current $J_N$; (c) the differential current $J_D$.
The dashed lines depict the linear approximations of the currents given by 
$J_{\alpha}^{\rm (lin)}=\sum_{\beta} L_{\alpha,\beta} A_{\beta}$ 
with the linear response coefficients (\ref{LEE})-(\ref{Lii}).}
\label{fig2}
\end{figure}

The three currents (\ref{JES})-(\ref{JD}) are depicted in Fig.~\ref{fig2} as a function of the affinity $A_D$ associated with the separation factor (\ref{f}).  They are compared with their linear approximations, $J_{\alpha}^{\rm (lin)}=\sum_{\beta} L_{\alpha,\beta} A_{\beta}$, given in terms of the linear response coefficients (\ref{LEE})-(\ref{Lii}).  We observe that the currents manifest important nonlinear behavior and that 
their linear approximation may soon become very crude away from equilibrium.   The differential current $J_D$ is observed in Fig. \ref{fig2}c to be positive already in a small interval at negative values of the corresponding affinity $A_D$.  It is in this small interval that mass separation occurs because the differential current $J_D>0$ is there opposite to the affinity $A_D<0$.

\subsection{Fluctuation theorem and entropy production}

Because of microreversibility, the fluctuating transfers of energy $\Delta E$ and particles
\bea
&& \Delta N \equiv N_{1}+N_{2} \, , \\
&& \Delta D \equiv N_{1}-N_{2} \, , 
\eea
obey the following fluctuation theorem:
\be
\frac{{\mathcal P}_t(\Delta E, \Delta N, \Delta D)}{{\mathcal P}_t(-\Delta E, -\Delta N, -\Delta D)} 
\simeq \exp(A_E\,  \Delta E + A_N \Delta N + A_D \Delta D) \qquad\mbox{for} \quad t\to\infty \, ,
\label{FTS}
\ee
and the response coefficients are related to the statistical cumulants by Eqs. (\ref{L})-(\ref{B}).
The thermodynamic entropy production (\ref{2ndlaw}) is here given by
\be
\frac{1}{k}\frac{d_{\rm i}S}{dt} = A_E\,  J_E + A_N\,  J_N +A_D\,  J_D \geq 0 \, ,
\label{entrprodS}
\ee
and is always non negative as the consequence of the fluctuation theorem.
We notice that this non-negativity is otherwise non trivial to establish, given the analytic expressions
of the currents (\ref{JES})-(\ref{JD}) and the affinities (\ref{AE}), (\ref{AN}), and (\ref{AD}).

\begin{figure}[htbp]
{\scalebox{0.4}{\includegraphics{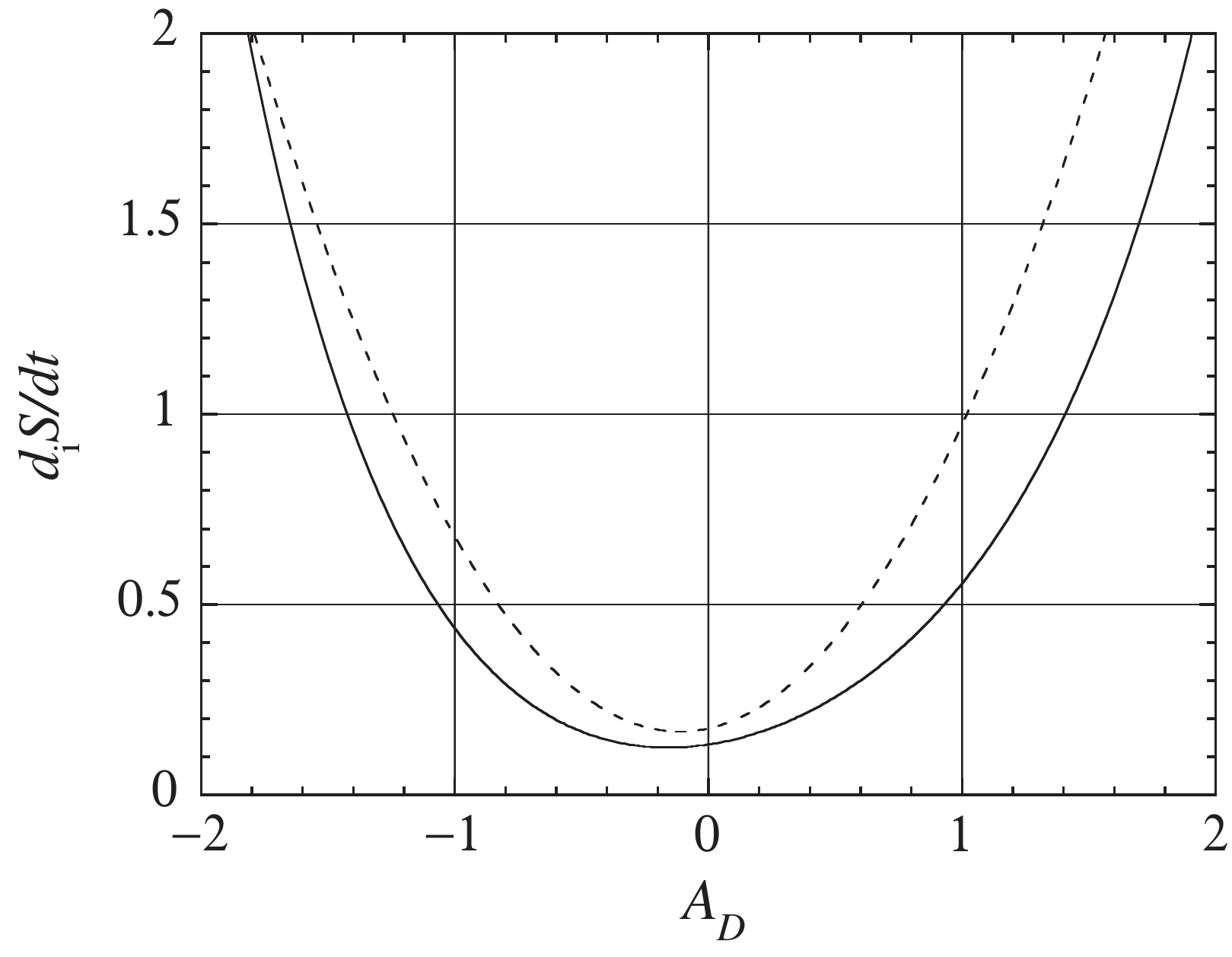}}}
 \caption{The thermodynamic entropy production (\ref{entrprodS}) versus the affiinity $A_D$ for effusion
 in the same conditions as in Fig.~\ref{fig2}. The dashed line depicts its approximation given by the
 linearized currents $J_{\alpha}^{\rm (lin)}=\sum_{\beta} L_{\alpha,\beta} A_{\beta}$ 
with the linear response coefficients (\ref{LEE})-(\ref{Lii}).}
\label{fig3}
\end{figure}

The positivity of the entropy production under nonequilibrium conditions is illustrated in Fig.~\ref{fig3} where we see that it is overestimated by its approximation based on the linearized currents.

\begin{figure}[htbp]
\begin{tabular}{cccccc}
  \rotatebox{0}{\scalebox{0.5}{\includegraphics{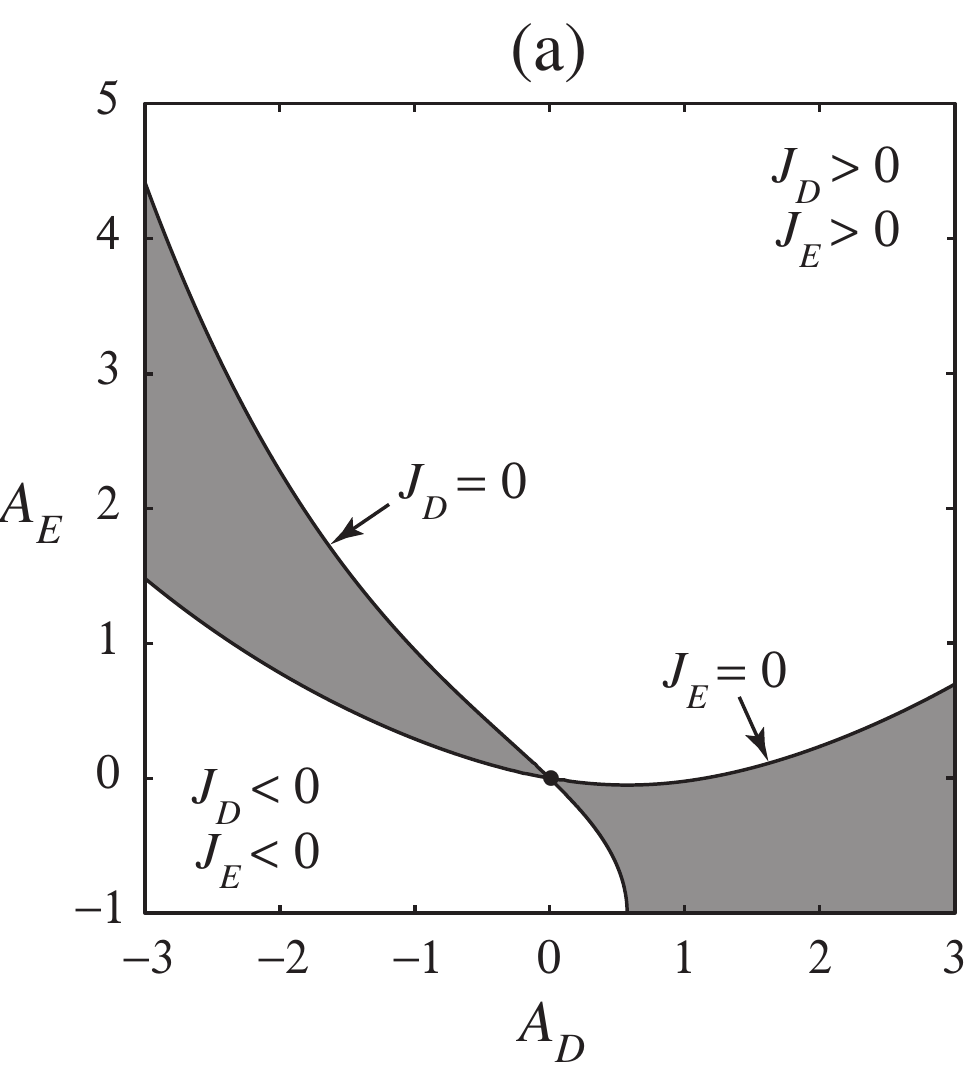}}}
  &&&&&
  \rotatebox{0}{\scalebox{0.5}{\includegraphics{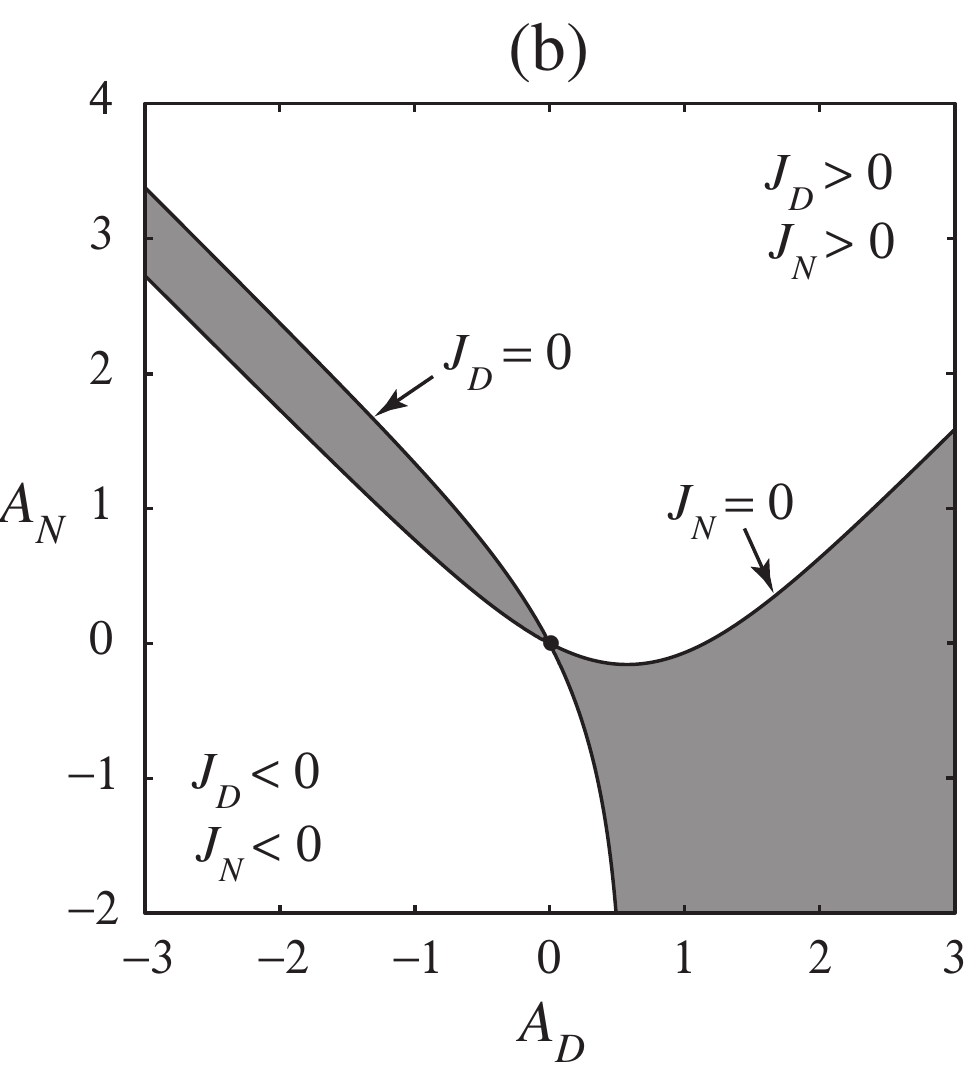}}} \\
\end{tabular}
\caption{Effusion of a binary mixture of particles of masses $m_1=1$ and $m_2=10$ at the temperature $kT=1$ and densities $n_1=n_2=1$ through a pore of area $\sigma=1$: (a) the plane $(A_D,A_E)$ at $A_N=0$; (b) the plane $(A_D,A_N)$ at $A_E=0$. The domains where the currents $J_D$ and $J_N$ have the same direction are in white
and those where the currents have opposite directions are in grey.}
\label{fig4}
\end{figure}

The non-negativity of the entropy production has implications in the organization of the domains where
the different currents take opposite directions in the three-dimensional space of the affinities.
Figure \ref{fig4} represents two sections in this space: the section at $A_N=0$ in Fig.~\ref{fig4}a and
the section at $A_E=0$ in Fig.~\ref{fig4}b.

In the plane $(A_D,A_E)$ depicted in Fig.~\ref{fig4}a, the directions of the currents $J_D$ and $J_E$ switch on the lines $J_D=0$ and $J_E=0$.  Both lines intersect at the state of thermodynamic equilibrium located at the origin $A_D=A_E=0$.  In the vicinity of this point, the linear approximation of the currents holds, but the lines $J_D=0$ and $J_E=0$ become curved away from equilibrium because of the nonlinear transport effects.
The second law of thermodynamics determines the signs of the currents in the different domains limited by the lines $J_D=0$ and $J_E=0$.  Indeed, the entropy production (\ref{entrprodS}) implies that $A_DJ_D\geq 0$ on
the line $J_E=0$ in the plane $A_N=0$, whereupon we find that $J_D>0$ for $A_D>0$, as it is indeed the case on the right-hand side of the origin in Fig.~\ref{fig4}a.  As a related consequence, the differential current has the opposite sign $J_D<0$ on the left-hand side where $A_D<0$, again along the line $J_E=0$.  A similar reasoning determines the sign of the current $J_E$ along the line $J_D=0$, as seen in Fig.~\ref{fig4}a.
Therefore, both currents $J_D$ and $J_E$ have the same sign in the white domains and they have opposite signs in the grey domains of Fig.~\ref{fig4}a, in consistency with the second law.

On the other hand, the entropy production (\ref{entrprodS}) reduces to $A_N J_N + A_D J_D \geq 0$
in the plane $A_E=0$ depicted in Fig.~\ref{fig4}b, where the lines corresponding to $J_D=0$ and $J_N=0$ are represented.  Here also, the different domains are organized around the thermodynamic equilibrium state at the origin according to the second law of thermodynamics.

\subsection{Entropy efficiency}

A concept of {\it entropy efficiency} has been introduced by Onsager \cite{O39,JF46}
as the efficiency of the thermodiffusive separation process to decrease Gibbs' mixing entropy
at the expense of increasing the other contributions of entropy.
In the case of effusion, the entropy efficiency can be defined as minus the ratio between the contribution to entropy production caused by the mixing of the two species over the other sources of dissipation:
\be
\eta \equiv - \frac{A_DJ_D}{A_EJ_E+A_NJ_N} \, .
\label{eff}
\ee
This quantity characterizes the efficiency of mass separation in terms of the power of the differential current divided by the power dissipated by the energy and the total particle currents.

In the regimes where mass separation is effective, the differential current $J_D$ is opposite to the corresponding affinity $A_D=\ln(1/\sqrt{f})$ so that $A_DJ_D<0$.  Therefore, the efficiency (\ref{eff}) is positive because
the non-negativity of the entropy production (\ref{entrprodS}) implies that $A_EJ_E+A_NJ_N \geq -A_DJ_D >0$.
Furthermore, the entropy efficiency (\ref{eff}) cannot exceed unity, which is its maximum value allowed by the second law of thermodynamics (\ref{entrprodS}):
\be
\eta \leq 1 \, ,
\label{max_eff}
\ee
in the regimes of effective mass separation where $\eta >0$.

With regard to the expressions (\ref{JES})-(\ref{JD}) of the currents, we notice that the entropy efficiency (\ref{eff}) depends only on four variables:
\be
\eta=\eta(kTA_E,A_N,A_D;q) \, ,
\ee
that are the combination $kTA_E=(T/T')-1$ of the temperature with the energy affinity, the affinities $A_N$ and $A_D$,
and, finally, the ratio of the effusion rates:
\be
q \equiv \frac{r_1}{r_2} = \frac{n_1}{n_2} \sqrt{\frac{m_2}{m_1}} \, ,
\label{q}
\ee
which plays a central role.  This result shows that the efficiency of mass separation by effusion depends on both the mass ratio and the initial concentration ratio.

In the linear regime close to the thermodynamic equilibrium where the currents can be replaced by their linear
approximation $J_{\alpha}^{\rm (lin)}=\sum_{\beta} L_{\alpha,\beta} A_{\beta}$ 
in terms of the linear response coefficients (\ref{LEE})-(\ref{Lii}), the maximum value of the entropy efficiency is given by
\be
\eta_{\rm max}^{\rm (lin)} = \left( \frac{\sqrt{q}-1}{\sqrt{q}+1}\right)^2 \, ,
\label{lin_max_eff}
\ee
which is reached at $A_E=0$ and $A_D=-A_N(\sqrt{q}-1)/(\sqrt{q}+1)$.
Accordingly, the maximum efficiency can approach the unit value if either $q\gg 1$ or $q\ll 1$.
However, we should expect deviations from the prediction (\ref{lin_max_eff}) in the nonlinear regimes further away from equilibrium, which is indeed the case.

\begin{figure}[htbp]
{\scalebox{0.37}{\includegraphics{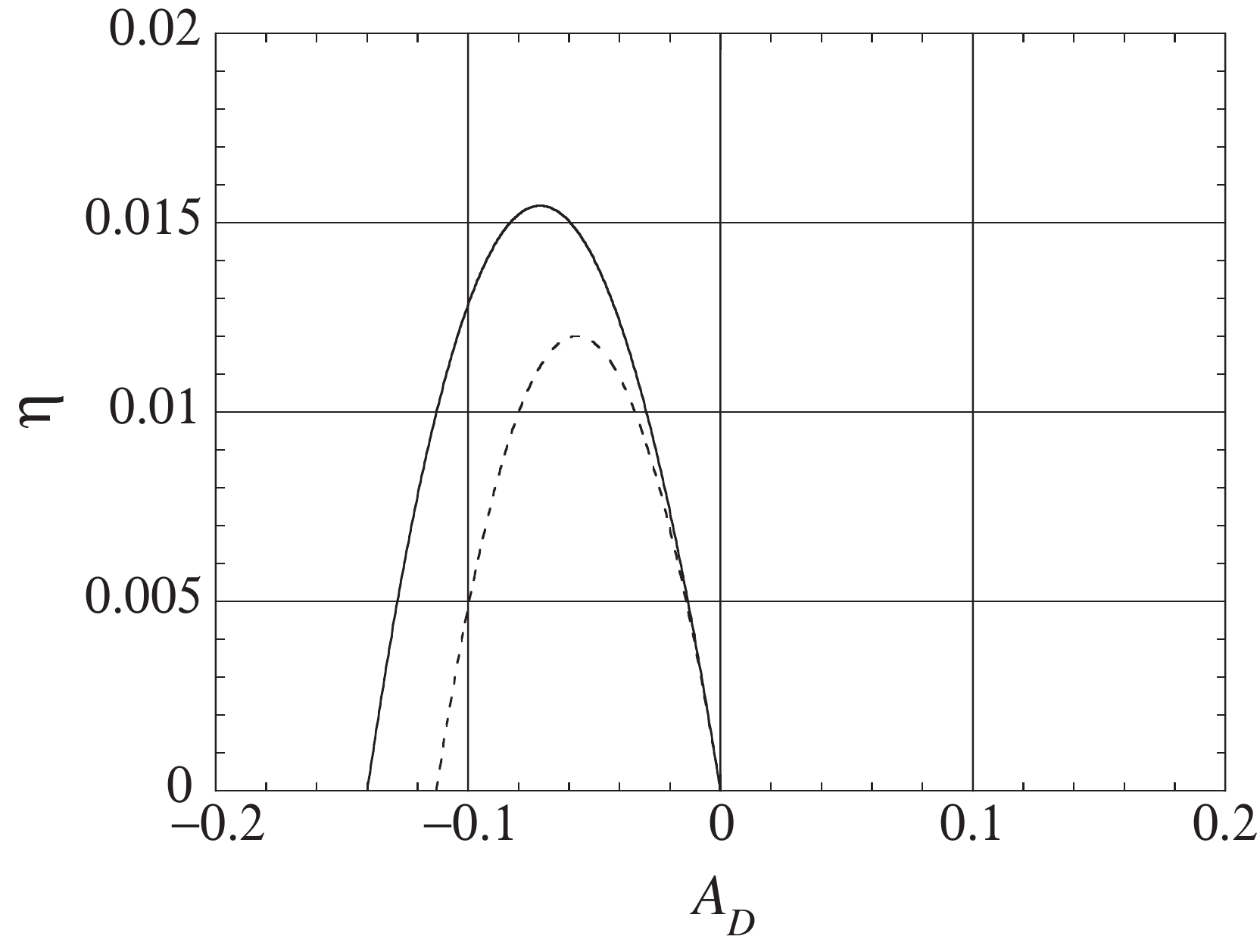}}}
\caption{The efficiency (\ref{eff}) versus the affinity $A_D$ under the same conditions as in Figs.~\ref{fig2} and~\ref{fig3},
for which $q=1.58$.
The dashed line depicts the efficiency for the linear approximation of the currents,
which would reach the maximum value $\eta_{\rm max}^{\rm (lin)} =0.013$ if $A_E=0$.}
\label{fig5}
\end{figure}

\begin{figure}[htbp]
{\scalebox{0.37}{\includegraphics{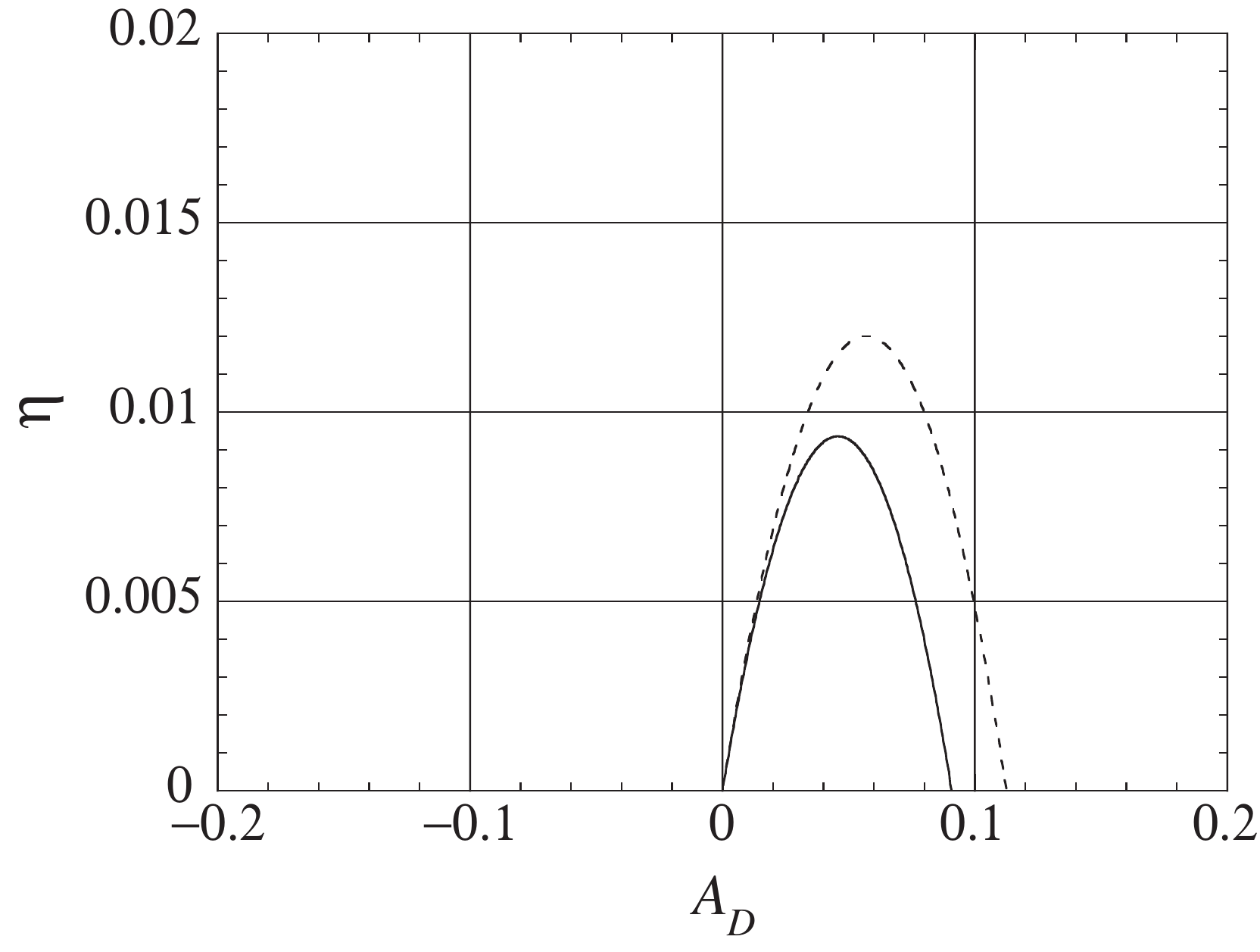}}}
\caption{The efficiency (\ref{eff}) versus the affinity $A_D$ for the effusion 
at the affinities $A_E=-0.1$ and $A_N=-0.3$
of a binary mixture of particles of masses $m_1=1$ and $m_2=10$
at the temperature $kT=1$ and densities
$n_1=1$ and $n_2=2$ in the reservoir $\mathcal{R}$
The pore is of area $\sigma=1$.  
The dashed line depicts the efficiency for the linear approximation of the currents.
Here also, $q=1.58$ and $\eta_{\rm max}^{\rm (lin)} =0.013$ if $A_E=0$.}
\label{fig6}
\end{figure}

Figure~\ref{fig5} depicts the efficiency (\ref{eff}) under the same conditions as in Figs.~\ref{fig2} and~\ref{fig3}.  As seen in Fig.~\ref{fig5}, the efficiency is positive in a small interval of values of the affinity $A_D$ extending from the locus where $J_D(A_D)=0$ to $A_D=0$.  It is in this small interval that mass separation is powered by the other currents $J_E$ and $J_N$. 
In Fig.~\ref{fig5}, the exact value of the efficiency is compared with the prediction of the linear approximation, which is
observed to underestimate the exact value.  However, if negative affinities $A_E$ and $A_N$ are taken, we observe in Fig.~\ref{fig6} that the linear approximation gives an overestimation of the actual efficiency.
Since the conditions of Figs.~\ref{fig5} and~\ref{fig6} are close to equilibrium and since the ratio (\ref{q}) is of the order of unity, the efficiency remains very small with respect to its maximum possible value (\ref{max_eff}). 

\begin{figure}[htbp]
{\scalebox{0.4}{\includegraphics{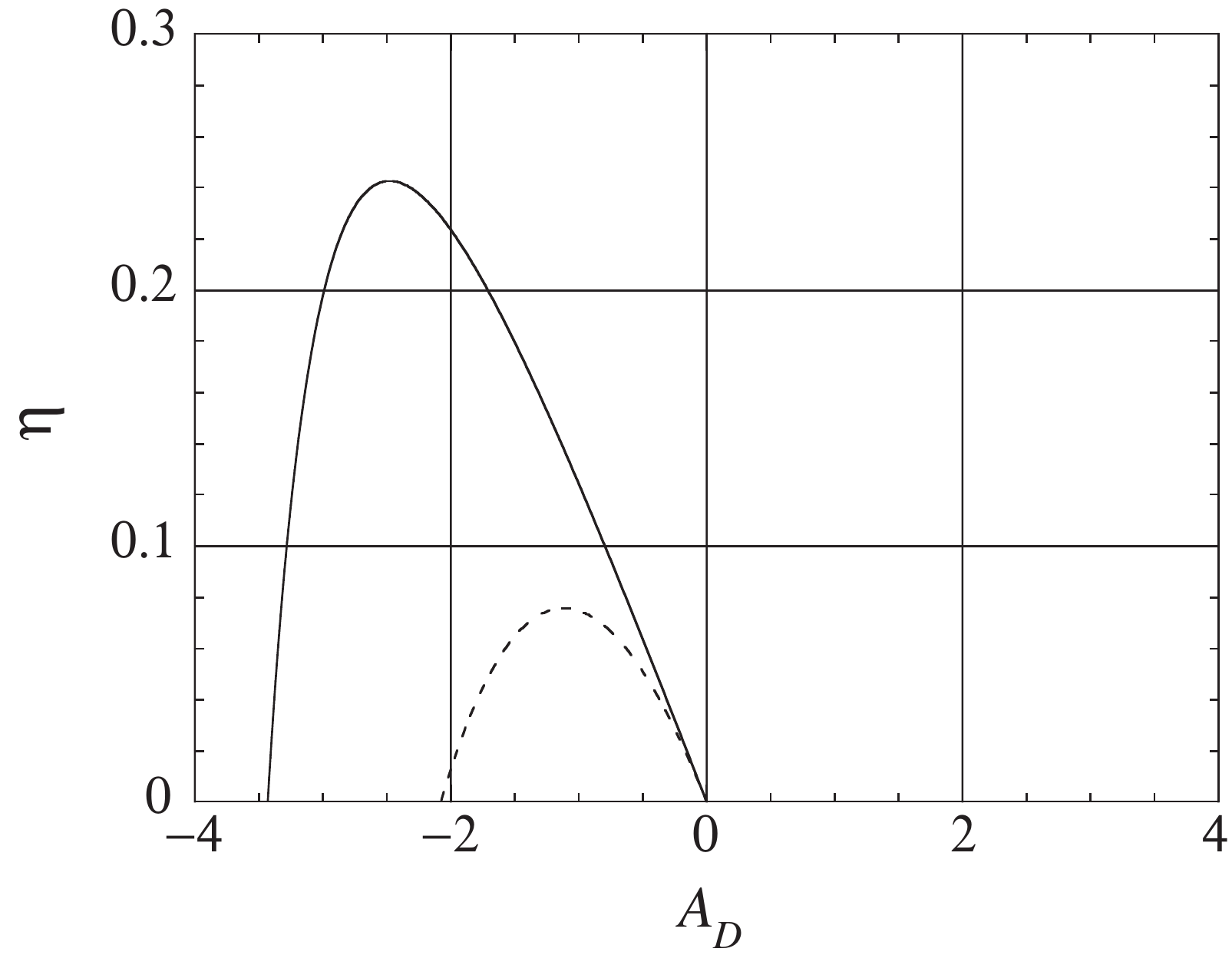}}}
\caption{The efficiency (\ref{eff}) versus the affinity $A_D$ for mixtures of masses $m_1=1$ and $m_2=10$.  The left reservoir $\mathcal{R}$ is at the temperature $kT=1$ and densities $n_1=n_2=1$ while the right reservoir $\mathcal{R}'$ is at the affinities $A_E=0.5$ and $A_N=3$. The ratio of the effusion rates is here given by $q=3.16$.  The dashed line depicts the efficiency for the linear approximation of the currents, which would reach the maximum value $\eta_{\rm max}^{\rm (lin)} =0.078$ if $A_E=0$.}
\label{fig7}
\end{figure}

Figure~\ref{fig7} shows the efficiency under conditions 
further away from equilibrium where the effect of nonlinearities is stronger.
The maximum value of the efficiency is here larger than in the conditions of Fig.~\ref{fig5}.
Moreover, the actual value of the efficiency turns out to be much larger than the value (\ref{lin_max_eff}) 
predicted by the linear approximation.  We remark that the actual value would be smaller than its linear approximation 
if the affinities $A_E$ and $A_N$ were negative, as already observed in Fig.~\ref{fig6}.  Accordingly, the largest values of the efficiency are typically obtained for positive rather than negative affinities $A_E$ and $A_N$.
These results clearly demonstrate that the nonlinear dependence of the currents on the affinities 
plays an important role in the process of mass separation by effusion.

\begin{figure}[htbp]
{\scalebox{0.5}{\includegraphics{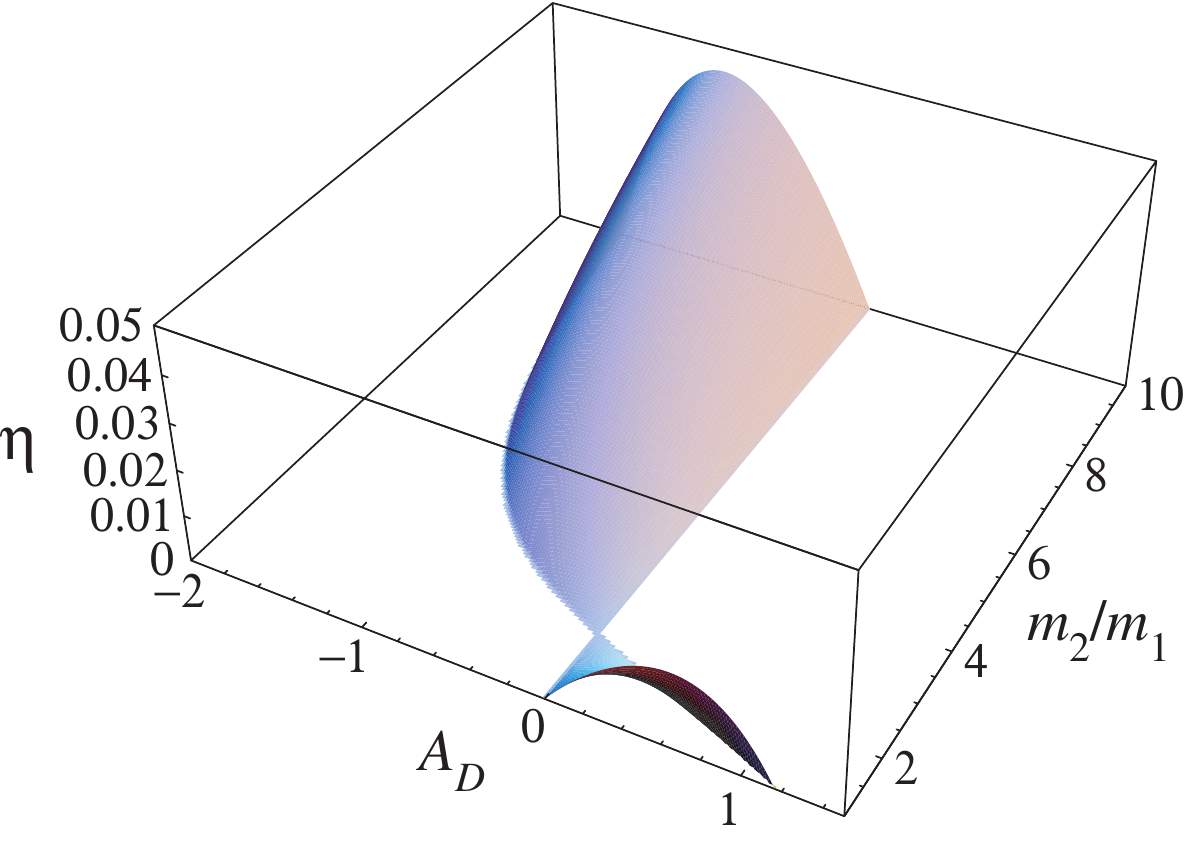}}}
 \caption{The entropy efficiency (\ref{eff}) for mass separation by effusion versus the affinity $A_D$ and the mass ratio $m_2/m_1$.  The left reservoir $\mathcal{R}$ is at the temperature $kT=1$ and densities $n_1=1$ and $n_2=1.5$, while the right reservoir $\mathcal{R}'$ is at the affinities $A_E=2$ and $A_N=0$.
 The entropy efficiency vanishes at the mass ratio $m_2/m_1=(n_2/n_1)^2=2.25$ on the straight line $A_D=0$.}
\label{fig8}
\end{figure}

\begin{figure}[htbp]
{\scalebox{0.5}{\includegraphics{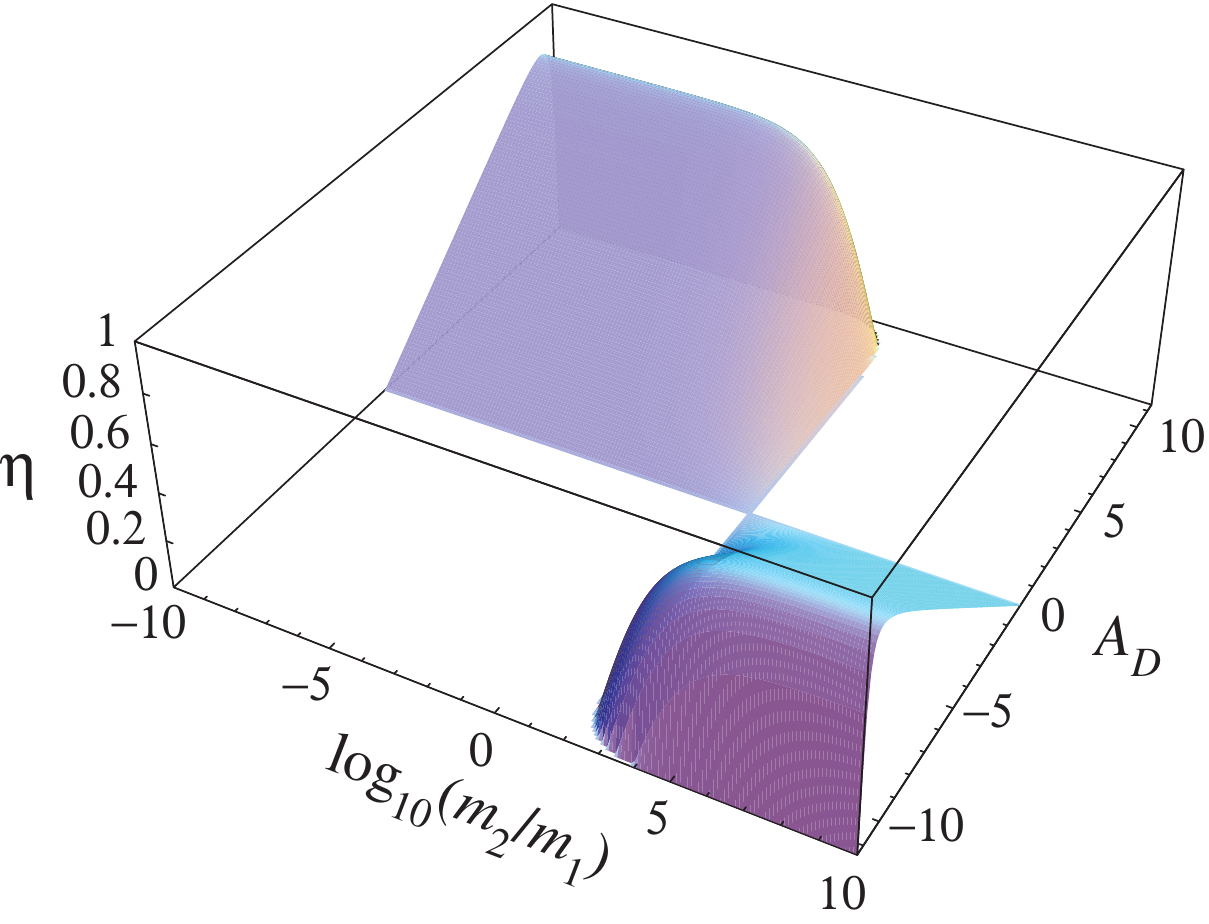}}}
  \caption{The entropy efficiency (\ref{eff}) for mass separation by effusion versus the affinity $A_D$ and the logarithm $\log_{10}(m_2/m_1)$ of the mass ratio, for the right reservoir $\mathcal{R}'$ at the affinities $A_E=1$ and $A_N=10$, while the left reservoir $\mathcal{R}$ is at the temperature $kT=1$ and densities $n_1=1$ and $n_2=10$.
 The entropy efficiency vanishes at the mass ratio $\log_{10}(m_2/m_1)=2\log_{10}(n_2/n_1)=2$ on the straight line $A_D=0$.  Note that the efficiency nearly reaches its maximum possible value (\ref{max_eff}) allowed by the second law under these conditions corresponding to a large difference of pressure between both reservoirs.}
\label{fig9}
\end{figure}

\begin{figure}[htbp]
{\scalebox{0.5}{\includegraphics{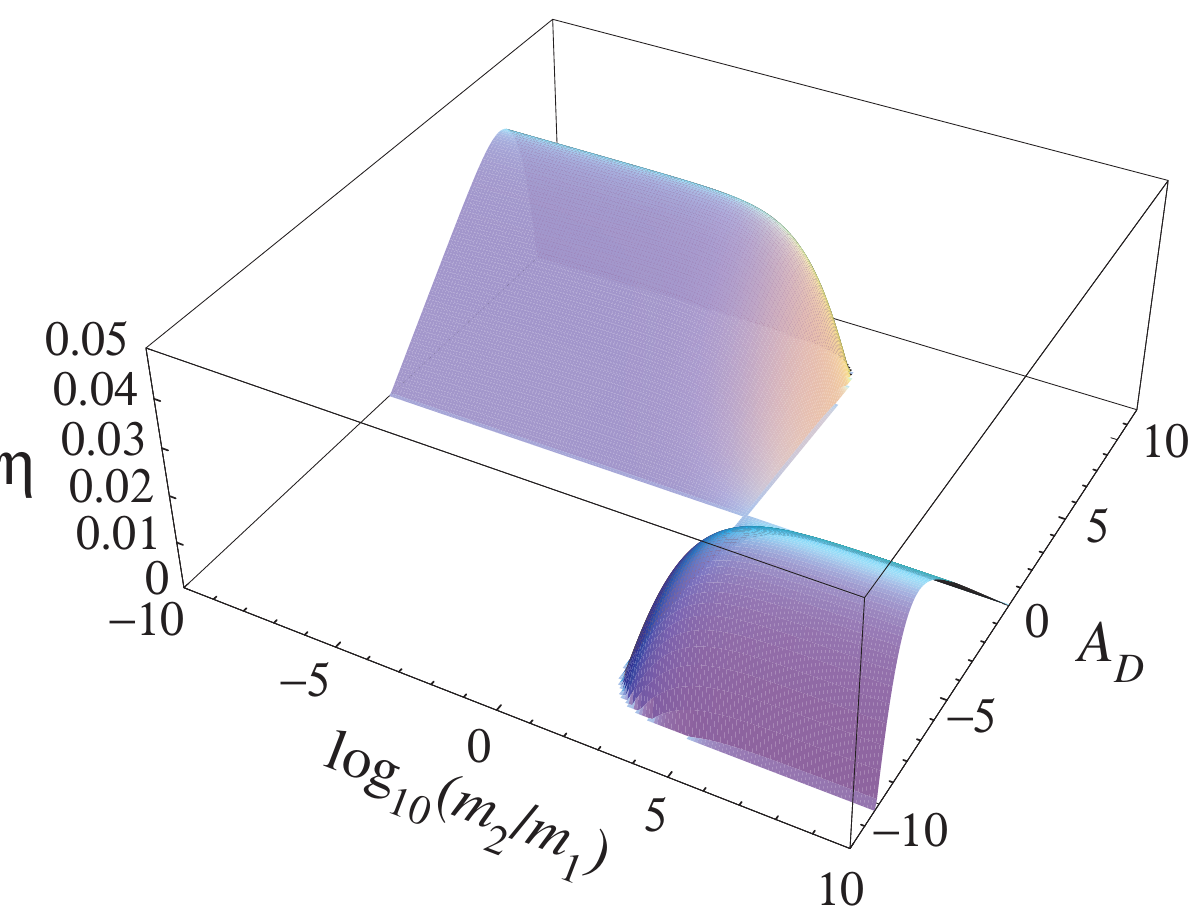}}}
 \caption{The entropy efficiency (\ref{eff}) for mass separation by effusion versus the affinity $A_D$ and the logarithm $\log_{10}(m_2/m_1)$ of the mass ratio, for the right reservoir $\mathcal{R}'$ at the affinities $A_E=100$ and $A_N=1$, while the left reservoir $\mathcal{R}$ is at the temperature $kT=1$ and densities $n_1=1$ and $n_2=10$.
 The entropy efficiency vanishes at the mass ratio $\log_{10}(m_2/m_1)=2\log_{10}(n_2/n_1)=2$ on the straight line $A_D=0$.  Note that the efficiency remains low under these conditions corresponding to a large difference of temperature between both reservoirs.}
\label{fig10}
\end{figure}

The dependence of the efficiency (\ref{eff}) on the mass ratio $m_2/m_1$ and the affinity $A_D$ is depicted in Figs.~\ref{fig8}-\ref{fig10} for different conditions.  In each plot, the efficiency is positive between the straight line $A_D=0$ and the curved line where $J_D=0$.  According to Eq.~(\ref{JD}) for the differential current, the intersection between the lines $A_D=0$ and $J_D=0$ happens if the ratio (\ref{q}) of the effusion rates takes the unit value $q=1$ while $A_N \neq -2\ln(1+kT\, A_E)$, in which case
\be
\frac{m_2}{m_1} = \left(\frac{n_2}{n_1}\right)^2 \, .
\label{q=1}
\ee

Figure~\ref{fig8} depicts the efficiency for a moderate temperature difference in a situation where the affinity $A_N$ is vanishing.  We observe that the efficiency is positive for $A_D<0$ if $q>1$ and for $A_D>0$ if $q<1$.
If $q=1$ so that the mass ratio is related by Eq. (\ref{q=1}) to the initial concentration ratio, mass separation cannot be performed by effusion.  Mass separation thus becomes effective for $q\neq 1$.
Nevertheless, the efficiency remains low under the conditions of Fig.~\ref{fig8}, which are close to equilibrium.

Figure~\ref{fig9} shows that the efficiency may approach its maximum possible value (\ref{max_eff}) allowed by the second law if the affinity $A_N$ takes large positive values, i.e., for very different pressures in both reservoirs.
Indeed, using the expressions (\ref{JES})-(\ref{JD}) for the currents and the definition (\ref{eff}), the maximum efficiency reached in this limit can be estimated to be given by
\be
\eta_{\rm max} \simeq \left\vert\frac{q-1}{q+1}\right\vert \qquad \mbox{for} \quad A_N \to +\infty \, ,
\ee
where $q$ is the ratio defined in Eq.~(\ref{q}).  We notice that this maximum efficiency is always larger than the maximum value (\ref{lin_max_eff}) of the linear regime.  Therefore, the efficiency can be much larger in these nonlinear regimes than predicted by the linear approximation.

In contrast, the efficiency cannot reach high values in the limit where the temperature difference is very large because the efficiency decreases as
\be
\eta_{\rm max} \simeq \left\vert\frac{q-1}{q+1}\right\vert \; \frac{\ln(kTA_E)}{kTA_E} \to 0  \qquad \mbox{for} \quad A_E \to +\infty \, ,
\ee
in this limit, as illustrated in Fig.~\ref{fig10}.

\section{Conclusions}
\label{sec:Conclusions}

In the present paper, we have reported a study of nonlinear transport properties in the process of mass separation
by effusion in the light of the recent advances in nonequilibrium statistical mechanics.
Effusion is the ballistic transport of particles through a small pore in a wall separating two reservoirs
containing gaseous mixtures at different temperatures, pressures, and concentrations.
This is the classical analogue of ballistic transport in open quantum systems 
where time-reversal symmetry relations known 
under the name of fluctuation theorem have been recently proved 
\cite{EHM09,AGMT09}.
For the effusion of gaseous mixtures, classical kinetic theory 
based on the Maxwellian velocity distribution can be used
to obtain the counting statistics and its generating function 
in terms of the affinities or thermodynamic forces, 
which are the control parameters
of the nonequilibrium constraints imposed 
by the particle reservoirs and driving the transport process.  

The fluctuation theorem for the currents established 
in Refs.~\cite{AG07JSP,A09} can be considered
for the stochastic processes of effusion.
This fluctuation theorem - which finds its origin in microreversibility - 
is the consequence of the relations (\ref{W-ratio})
between the forward and backward transition rates
and the affinities.  The main point is that the fluctuation theorem for the currents
implies that the nonlinear response coefficients 
obey remarkable relationships that are the generalizations
of the Onsager reciprocity relations \cite{AG04,AG06,AG07JSM},
as we have here demonstrated for effusion in gaseous mixtures.  
On the one hand, these relationships reveal that the nonlinear response coefficients
can be obtained from the statistical cumulants characterizing the fluctuations of the currents
in nonequilibrium steady states.  On the other hand, they establish symmetries among the coefficients
as given by Eqs.~(\ref{BEEEE})-(\ref{Biiii}) and Eqs.~(\ref{symNEEEi})-(\ref{symNEiii}).

In section \ref{sec:MS}, these considerations are applied to mass separation by effusion in binary mixtures
where the flows of two particle species are coupled to the energy flow.  
Here, we have introduced the total particle current and the differential current that are coupled
not only to the energy current but also together, allowing the differential current to be controlled
by both the energy and total particle currents.  

The nonlinear dependence of the three currents on the
three conjugated affinities is evidenced by comparing their actual values to their linear approximations.
These results show the importance to go beyond linear response theory in considering
the nonequilibrium thermodynamics of mass separation by effusion.
Because of the nonlinear transport effects, the entropy efficiency may be significantly larger than
it would be the case in the linear approximation.  

The second law of thermodynamics, which is here proved thanks to the fluctuation theorem,
imposes an upper bound on the entropy efficiency of mass separation by effusion.
This fundamental limit is not reached under moderate nonequilibrium conditions,
but can nevertheless be approached for large pressure differences between both reservoirs.

In conclusion, we see that our knowledge of transport phenomena can nowadays be extended
from the linear to the nonlinear regimes thanks to the recent advances in nonequilibrium statistical mechanics.

\begin{acknowledgments}
D.~Andrieux thanks the F.R.S.-FNRS Belgium for financial support.
This research is financially supported by the Belgian Federal Government
(IAP project ``NOSY").
\end{acknowledgments}



\begin{thebibliography}{99}
\bibitem{inequality}  This inequality is the consequence of Jensen's inequality $\left\langle f(X)\right\rangle \geq f(\langle X\rangle)$ holding for any convex function $f(X)$ and any expectation value $\langle\cdot\rangle$ with respect to some normalized probability distribution such that $\langle 1 \rangle =1$ \cite{GR80}.  
Here, the convex function is taken as $f(X)=\exp(-X)$.  
In the context of equilibrium statistical mechanics, this inequality is often 
called the Gibbs-Bogoliubov-Feynman bound \cite{C87}.
\bibitem{footnote} We notice that we could alternatively introduce the total mass current $J_M\equiv m_1 J_1+m_2J_2$ or the total current $J_G\equiv \sqrt{m_1} J_1+\sqrt{m_2}J_2=P/\sqrt{kT}-P'/\sqrt{kT'}$, which makes explicit Graham's law.
\end{thebibliography}
\end{document}